\documentclass[twocolumn,twocolappendix]{aastex631}

\usepackage{amsmath}
\usepackage{enumitem}

\begin{document}

\title{Exploring Realistic Nanohertz Gravitational-Wave Backgrounds}

\correspondingauthor{Bence B\'ecsy}
\email{bencebecsy@montana.edu}

\author[0000-0003-0909-5563]{Bence B\'ecsy}
\affiliation{%
 eXtreme Gravity Institute, Department of Physics, Montana State University, Bozeman, MT 59717, USA
}%
\affiliation{%
Department of Physics, Oregon State University, Corvallis, OR 97331, USA
}

\author[0000-0002-7435-0869]{Neil J.~Cornish}
\affiliation{%
 eXtreme Gravity Institute, Department of Physics, Montana State University, Bozeman, MT 59717, USA
}%

\author[0000-0002-6625-6450]{Luke Zoltan Kelley}
\affiliation{
 CIERA (Center For Interdisciplinary Exploration And Research In Astrophysics), Evanston, IL 60201, USA
}%
\affiliation{
Department of Astronomy, University of California at Berkeley, Berkeley, CA 94720, USA
}%



\begin{abstract}

Hundreds of millions of supermassive black hole binaries are expected to contribute to the gravitational-wave signal in the nanohertz frequency band. Their signal is often approximated either as an isotropic Gaussian stochastic background with a power-law spectrum, or as an individual source corresponding to the brightest binary. In reality, the signal is best described as a combination of a stochastic background and a few of the brightest binaries modeled individually. We present a method that uses this approach to efficiently create realistic pulsar timing array datasets using synthetic catalogs of binaries based on the Illustris cosmological hydrodynamic simulation. We explore three different properties of such realistic backgrounds which could help distinguish them from those formed in the early universe: i) their characteristic strain spectrum; ii) their statistical isotropy; and iii) the variance of their spatial correlations. We also investigate how the presence of confusion noise from a stochastic background affects detection prospects of individual binaries. We calculate signal-to-noise ratios of the brightest binaries in different realizations for a simulated pulsar timing array based on the NANOGrav 12.5-year dataset extended to a time span of 15 years. We find that $\sim6$\% of the realizations produce systems with signal-to-noise ratios larger than 5, suggesting that individual systems might soon be detected (the fraction increases to $\sim41$\% at 20 years). These can be taken as a pessimistic prediction for the upcoming NANOGrav 15-year dataset, since it does not include the effect of potentially improved timing solutions and newly added pulsars.
\end{abstract}



\section{\label{sec:intro}Introduction}

Pulsar timing arrays (PTAs) probe gravitational waves (GWs) with frequencies between a few and a few hundred nanohertz (nHz) by continuously monitoring millisecond pulsars (for a review see e.g.~\citealt{PTA_review, SteveBook}). The NANOGrav collaboration recently found evidence for a low-frequency stochastic process common to all pulsars in their 12.5-year dataset \citep{NANOGrav_12p5_gwb}. The three other PTAs also all found evidence for such a common red noise process (International Pulsar Timing Array (IPTA) \citealt{IPTA_DR2_GWB} , EPTA \citealt{EPTA_dr2_gwb}, PPTA \citealt{PPTA_dr2_gwb}). It is possible that these results are the first signs of a stochastic gravitational-wave background (GWB), but at this point none of the above analyses found strong evidence for Hellings-Downs (HD, \citealt{HD}) spatial correlations that are the tell-tale sign of a gravitational background. However, this is not surprising given that the power in cross-correlations is about an order of magnitude less then the total power of a GWB. This means that we can expect to detect a common process before we detect the HD correlations between pulsars (see e.g.~\citealt{Astro4Cast, romano_crn_vs_gwb}). Upcoming PTA datasets will decide whether this red noise process is the GWB or not.

Once a firm detection of the GWB is established, the next task will be to identify its origin. The theoretically most favored source of a GWB at nHz frequencies is an ensemble of inspiralling supermassive black hole binaries (SMBHBs, see e.g.~\citealt{smbh_gwb_rajagopal, smbh_gwb_sesana, smbh_gwb_enoki, Luke_paper2}). However, there are several other proposed processes that can form such a low-frequency GWB, like cosmological phase transitions \citep{NANOGrav_12p5_phase_transition}, cosmic strings \citep{EPTA_cosmic_string, KaiSchmitz_cosmic_stings}, primordial black holes \citep{primordial_bh_gwb}, inflation \citep{inflation_gwb}, etc. These models all have different predictions for the spectral shape of the common signal, which presents the possibility of distinguishing these models \citep{andrew_multiple_gwbs}. However, definitively identifying the source of the GWB based solely on its spectrum will be challenging given the large uncertainty in both measurement \citep{Astro4Cast} and model predictions. Thus any additional predictions of these models beyond their spectra could be useful in determining the source of the GWB.

In this paper we explore various properties of realistic GWBs from SMBHBs, some of which might help distinguish that scenario from others. As we will see, most of these properties rely on the fact that an SMBHB-based GWB is built from a finite number of sources. The statistics of individual sources in a realistic simulation were first investigated in \citet{SVV09}, while \citet{rosado_expected_properties} calculated detection probabilities of individual sources in the presence of a GWB. The general question of when the signal from a finite collection of individual sources becomes effectively stochastic was investigated in \citet{PhysRevD.92.042001}, while
the effects of the finite population on the detection prospects of the GWB have been studied in \citet{Neil_Laura_finite_number}. Here we are instead focusing on how these effects can help us discern the origin of the GWB. Our simulations are using synthetic SMBHB catalogs based on the Illustis cosmological simulation \citep{Luke_single_source}. We introduce and validate a method that can rapidly produce simulated PTA datasets from millions of SMBHBs by modeling the large majority of the sources as a GWB and directly simulating the few brightest binaries in each frequency bin (see Section \ref{sec:methods}). We use this simulation method to investigate how the central limit theorem breaks down for such an SMBHB population (see Section \ref{ssec:spectrum}), how the finite number of sources can lead to anisotropic GWBs (see Section \ref{ssec:isotropy}), and how the variance of the HD correlations is affected by the properties of the SMBHB population (see Section \ref{ssec:hd-variance}). We also explore the detectability of individual binaries in the presence of a GWB by calculating the distribution of signal-to-noise ratios (SNRs) of the brightest binaries in multiple realizations (see Section \ref{sec:detectability}). We summarize and offer concluding remarks in Section \ref{sec:conclusion}.

\section{\label{sec:methods}Simulation methods}

\subsection{\label{ssec:smbh_pop}SMBH population model}

Our simulations are based on synthetic catalogs of the SMBHB population representative of the entire observable universe. These are derived from the Illustris cosmological hydrodynamic simulations (see e.g.~\citealt{Illustris}), as implemented in the \texttt{holodeck} code \citep{holodeck_methods}, by modeling small-scale astrophysical processes in a post-processing step. This treatment accounts for environmental effects relevant to binary hardening like dynamical friction, stellar scattering, drag from a circumbinary disk, and GW emission (for more details see \citealt{Luke_paper1, Luke_paper2, Luke_single_source}).  The collection of mergers observed in the Illustris simulation volume can be resampled multiple times to create new realizations of a simulated catalog of SMBHBs for the entire past light-cone of the observer. The standard free parameter in this implementation of SMBH binary populations is the binary lifetime.  Additionally, to produce a GWB amplitude roughly consistent with the observed common process in the NANOGrav 12.5-year dataset \citep{NANOGrav_12p5_gwb}, we have tuned the volume density of mergers and the distribution of SMBH masses.

\begin{figure}[htb]
 \centering
   \includegraphics[width=\columnwidth]{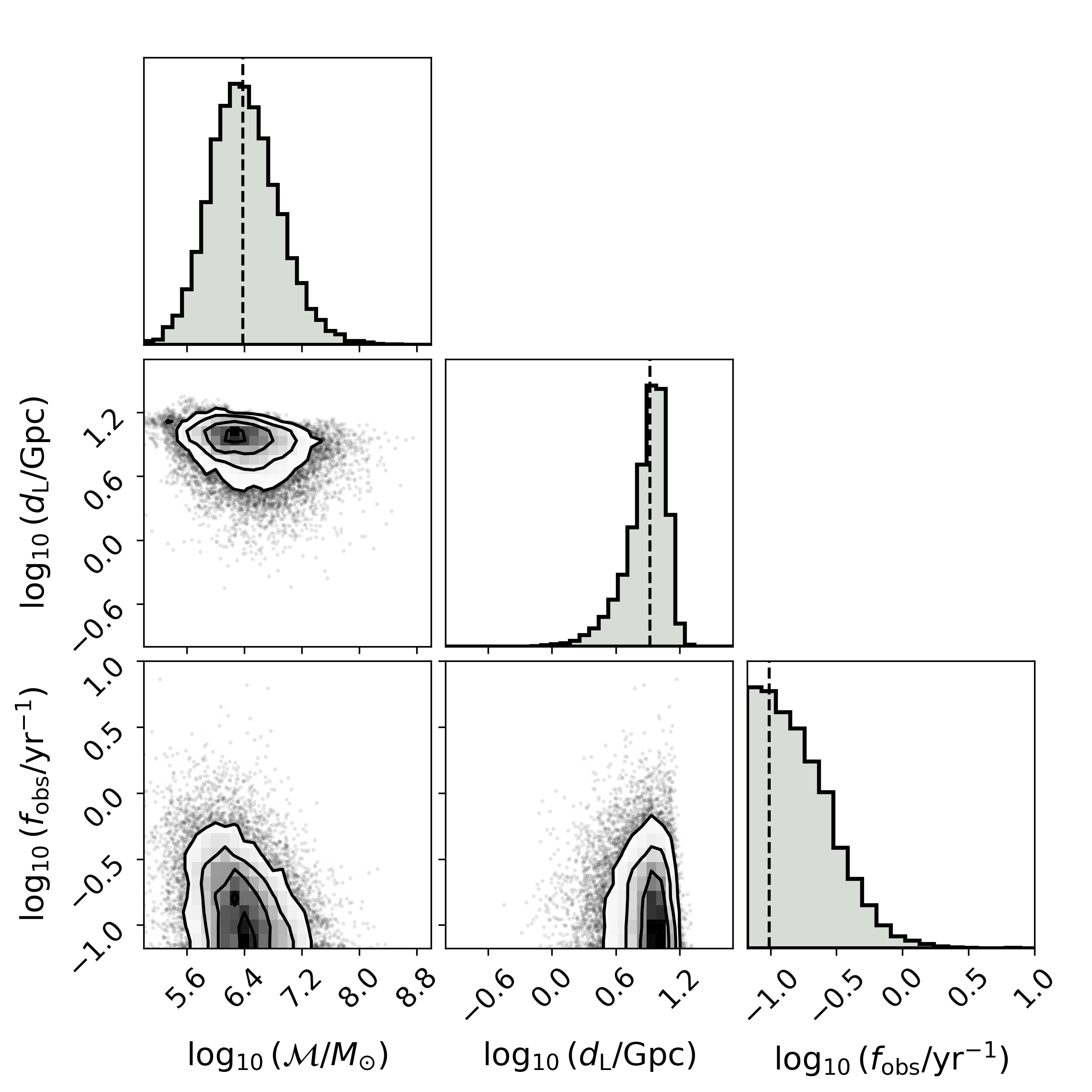}
 \caption{Distribution of binary parameters in the simulated SMBHB catalog we use. The distribution was artificially cut off below observer-frame GW frequencies of (15 year)$^{-1}$, which would not be observable with the 15 year observing timespan considered here. Vertical dashed lines indicate the median value of each parameter.}
 \label{fig:dataset_corner_plot}
\end{figure}

Figure \ref{fig:dataset_corner_plot} shows the distribution of the source-frame chirp mass ($\mathcal{M}$), the luminosity distance ($d_{\rm L}$), and the observer-frame GW frequency ($f_{\rm obs}$) of such a simulated dataset of SMBHBs. Here we applied a lower frequency cutoff at (15 year)$^{-1}$, corresponding to the observational timespan we consider in this paper and the observational timespan of the upcoming NANOGrav 15-year dataset \citep{nanograv_15_data}. This particular realization has about 115 million binaries. Note that the number density is dominated by low-frequency, low-mass, faraway systems.

\subsection{\label{ssec:split_method}Isotropic GWB plus bright binaries}

The timing residuals in a PTA dataset at observing epochs $t_i$ can be written as:
\begin{equation}
    r(t_i) = r_{\rm n} (t_i)  + r_{\rm GW} (t_i),
\end{equation}
where $r_{\rm n}$ describes residuals due to every noise source, while $r_{\rm GW}$ is the contribution of GWs. In the case of a population of $N$ SMBHBs producing GWs in the observable universe, one can write:
\begin{equation}
    r_{\rm GW} (t_i) = \sum_{j=1}^N s_{j} (t_i; \boldsymbol\theta_j),
\label{eq:direct_cw}
\end{equation}
where $s_{j}$ is the timing residual response to the GW signal from the $j$th binary, which is described by the parameters:
\begin{multline}
\boldsymbol\theta_j = \{ \mathcal{M}, d_{\rm L}, f_{\rm obs}, \theta, \phi, \iota, \psi , \\ \Phi_0 , \Phi_1, \dots, \Phi_{N_{\rm PSR}}, L_1, \dots, L_{N_{\rm PSR}}\},
\end{multline}
where $\theta$ and $\phi$ parametrize the location of the source on the sky, $\iota$ is the inclination of the binary's orbit, $\psi$ is the polarization angle, $\Phi_0$ ($\Phi_i$) is the initial phase of the GW signal at the Earth (at the $i$th pulsar), $L_i$ is the distance to the $i$th pulsar, and $N_{\rm PSR}$ is the number of pulsars (for an explicit expression of $s_{j} (t_i; \boldsymbol\theta_j)$ see e.g.~Eq.~(10) in \citealt{nanograv_11_cw}). While Eq.~(\ref{eq:direct_cw}) is a valid description of the total GW signal, it is impractical both for data analysis and for simulating signals. In terms of data analysis, this model has $(8+N_{\rm PSR}) N + N_{\rm PSR}$ parameters\footnote{The pulsar distance parameters are the same for each GW source, but the pulsar phase parameters introduced to help convergence will be different for each binary.}, which are practically impossible to explore in the realistic case, where $N \sim \mathcal{O}(10^6-10^9)$. In terms of simulating PTA datasets, individually calculating the response of millions of binaries becomes a computationally expensive task.

We will show that both of these problems can be averted by expressing the total contribution to the residuals as:
\begin{equation}
 r_{\rm GW} (t_i) = r_{\rm GWB} (t_i, h_c^{\rm GWB}) + \sum_{k=1}^{N_{\rm freq}} \sum_{j=1}^{M} s_{j}^{(k)} (t_i; \boldsymbol\theta_j),
\label{eq:split_method}
\end{equation}
where $r_{\rm GWB}$ is an isotropic Gaussian stochastic background with the characteristic strain spectrum $h_c^{\rm GWB}$, $N_{\rm freq}$ is the number of frequency bins considered, and $s_{j}^{(k)}$ is the binary with the $j$th largest characteristic strain in the $k$th frequency bin. The second term in Eq.~(\ref{eq:split_method}) loops over all $N_{\rm freq}$ frequency bins and directly adds the contribution of the $M$ brightest sources in each bin. We set $h_c^{\rm GWB}$ in each frequency bin based on the remaining SMBHBs as:
\begin{equation}
 h_c^{\rm GWB} (f_k) = \sqrt{ \sum_{j=M+1}^{N_k} \left[ (h_c)_{j}^{(k)} \right]^2 },
\label{eq:gwb_sqsum}
\end{equation}
where $N_k$ is the total number of binaries in the $k$th frequency bin, $f_k$ is the central frequency of that bin, and 
\begin{equation}
 (h_c)_{j}^{(k)} = h_{j}^{(k)} \sqrt{f_j^{(k)} T_{\rm obs} }
\label{eq:h_c}
\end{equation}
is the characteristic strain of the $j$th brightest source in the $k$th frequency bin, $T_{\rm obs}$ is the total observation time, $f_j^{(k)}$ is the observer-frame GW frequency of that source, and $h_{j}^{(k)}$ is the sky and polarization averaged GW amplitude \citep{smbh_gwb_sesana}:
\begin{equation}
 h_{j}^{(k)} = \frac{8}{\sqrt{10}} \frac{(G \mathcal{M}_{\rm obs})^{5/3} (\pi f_{\rm obs})^{2/3}}{c^4 d_{\rm L}},
\end{equation}
where $\mathcal{M}_{\rm obs} = (1+z) \mathcal{M}$ is the observer-frame chirp mass. Note that Eq.~(\ref{eq:gwb_sqsum}) determines $h_c^{\rm GWB}$ by summing up the square of the characteristic strain for all binaries except the ones directly modeled in Eq.~(\ref{eq:split_method}). This is justified by the fact that when summing up sine-waves of a given frequency with random phase offsets, the expectation value of the total amplitude squared is given by the sum of the squares of individual amplitudes. This approximately still holds in this scenario, where the frequencies of sources in the same frequency bin is approximately the same.

\begin{figure}[htb]
 \centering
   \includegraphics[width=\columnwidth]{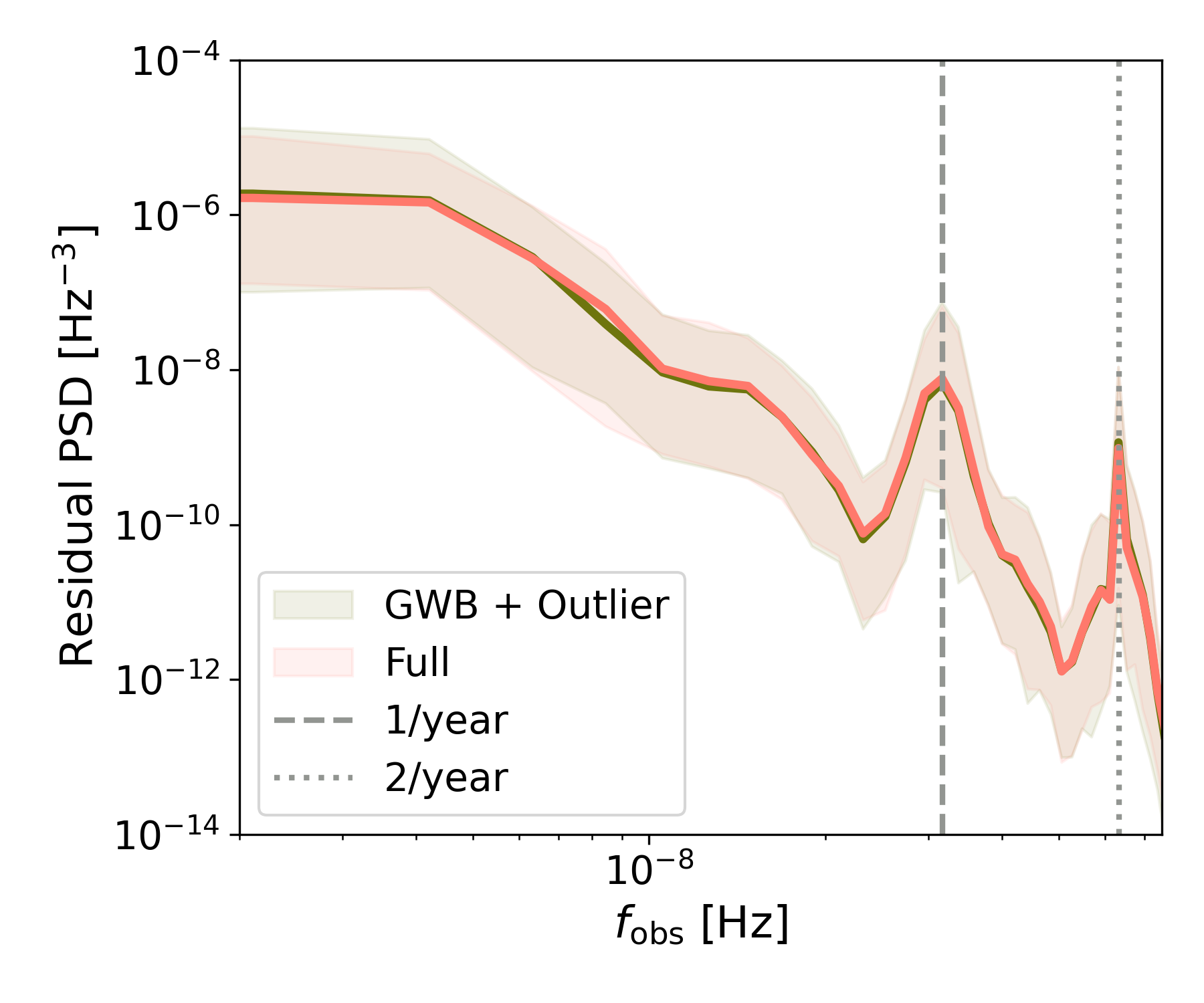}
 \caption{PSD of post-fit residuals for datasets where we individually simulated all binaries (red) and where we simulated a stochastic background and one thousand SMBHBs in each frequency bin (green). We show the median and 90\% credible intervals over 200 realizations as solid lines and shaded regions, respectively. The more efficient outlier method correctly reproduces both the median and the variance of the PSD from the full simulations.}
 \label{fig:split_method_validation}
\end{figure}

This method of describing a realistic background by an idealized isotropic Gaussian GWB and a sum of the few brightest sources can be used for both data analysis and simulation. The \texttt{BayesHopper} \footnote{Publicly available at: \url{https://github.com/bencebecsy/BayesHopper}} algorithm uses this approach to search for multiple individual SMBHBs in the presence of a GWB \citep{BayesHopper}, while in this paper we use this approach to efficiently create realistic simulated datasets. Note that for data analysis purposes, the ideal number of SMBHBs to model individually is dictated by Bayesian parsimony. However, for simulating signals we want to make sure that all the relevant details are captured, so we add more binaries directly than strictly necessary, and than what can be picked up by data analysis. To validate the use of this description for rapidly simulating realistic GWBs, we simulated 200 realizations of a GWB by directly simulating all binaries (as in Eq.~(\ref{eq:direct_cw})) and also by simulating a GWB and 1000 outlier SMBHBs in each frequency bin (as in Eq.~(\ref{eq:split_method})). Figure \ref{fig:split_method_validation} shows the median power spectral density (PSD) of post-fit residuals in a given simulated pulsar for both methods. The 90\% credible intervals are shown as shaded bands. Here we used a simulated noise-free dataset with a single pulsar, and evenly spaced observations to enable the use of fast Fourier transforms. We applied a Tukey window on the residuals before calculating the PSD to avoid spectral leakage. Both methods produce spectra with a significant variance over realizations, but both their median values and 90\% credible intervals show good agreement. Note that the two peaks correspond to frequencies of 1/year and 2/year, where the timing model fit introduces extra power. The agreement between the PSDs produced by the two methods show that they result in statistically equivalent frequency content. In Sections \ref{ssec:spectrum} and \ref{ssec:isotropy} we show that after removing the brightest binaries, the remaining signal satisfies the assumption of Gaussianity and statistical isotropy as well.

\section{\label{sec:background_properties}Properties of a realistic GW background}

In this section we explore how various characteristics of the background are affected by the fact that there is a finite number of sources contributing to it. We investigate the spectrum of the GWB in Section \ref{ssec:spectrum}, we test for statistical isotropy in Section \ref{ssec:isotropy}, and quantify the variance of the HD correlations in Section \ref{ssec:hd-variance}.

\subsection{\label{ssec:spectrum}GW spectrum}

\begin{figure}[htb]
 \centering
   \includegraphics[width=\columnwidth]{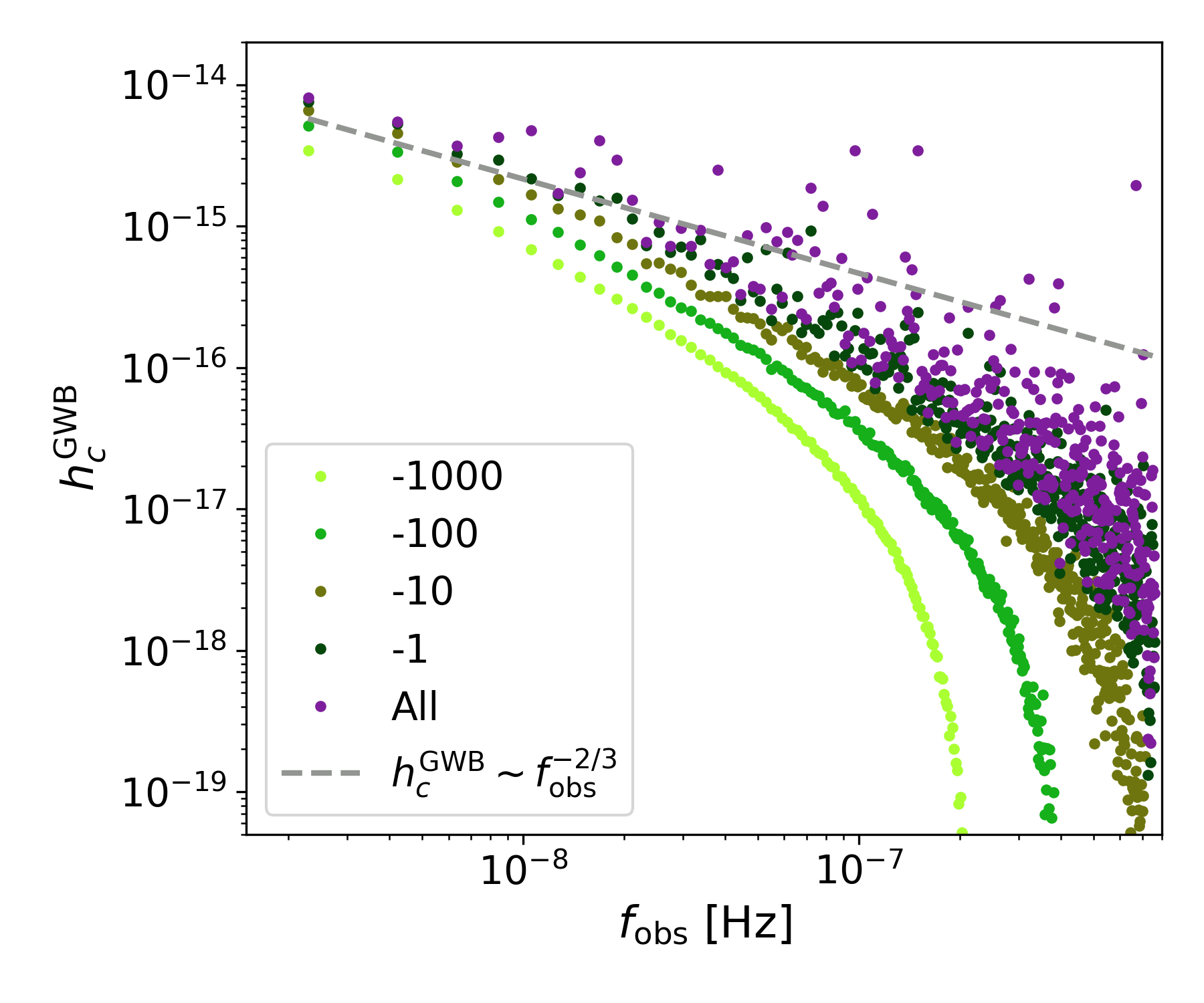}
 \caption{Characteristic strain spectrum for all binaries (purple) and for various numbers of the brightest binaries removed (different shades of green). We also show a power-law spectrum for reference (dashed gray line).}
 \label{fig:spectrum_variance}
\end{figure}

In the canonical description of a GWB arising from an infinite number of circular SMBHBs evolving purely through GW emission, the characteristic strain spectrum can be described as a power-law with a spectral index of -2/3, i.e.~$h_c^{\rm GWB} \sim f_{\rm obs}^{-2/3}$ \citep{Phinney}. The spectrum arising from a population of SMBHBs differs from this simple spectrum (see purple dots and dashed line in Fig.~\ref{fig:spectrum_variance}). The difference is most striking at high frequencies, where the population-based spectrum has a lower amplitude than the power-law model. This is due to the fact that the power-law model counts non-physical contributions from fractional sources \citep{Sesana_high_f_discrepancy}. The population-based spectrum also exhibits significant fluctuations at high frequencies. Fig.~\ref{fig:spectrum_variance} also shows the spectrum when we exclude the brightest 1/10/100/1000 binaries in each frequency bin \footnote{This corresponds to setting $M=1/10/100/1000$ in Eq.~(\ref{eq:gwb_sqsum})}. We can see that as we exclude more binaries, the scatter in the spectrum decreases. This suggests that by setting $M\gtrsim100$, the resulting spectrum is not dominated by a few bright binaries anymore. Note that as we remove more binaries, the spectra in Fig.~\ref{fig:spectrum_variance} terminate at lower frequencies, because at higher frequencies we removed all the binaries.

\begin{figure}[htb]
 \centering
   \includegraphics[width=\columnwidth]{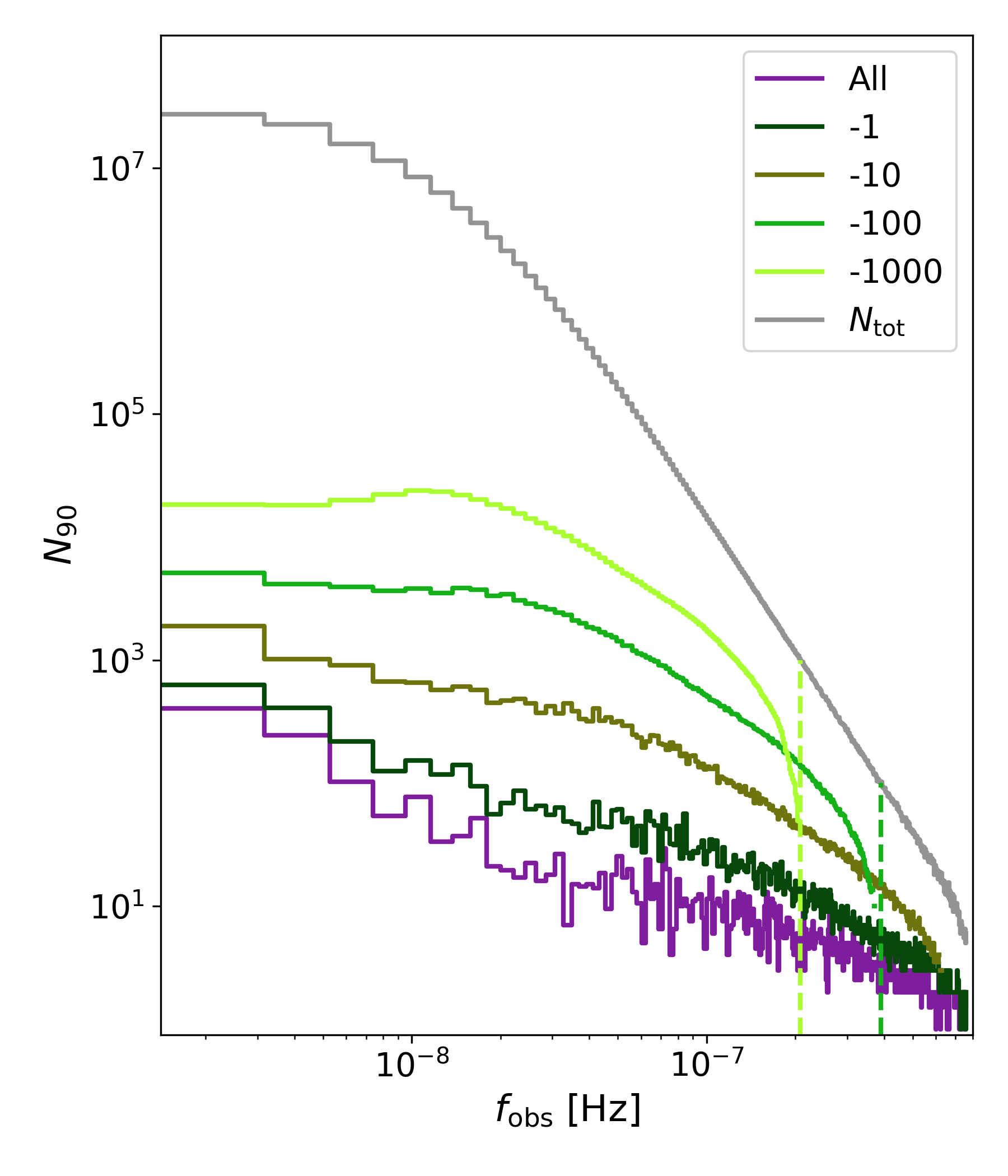}
 \caption{Number of binaries contributing 90\% of the total strain in each frequency bin ($N_{90}$). We show this for the whole dataset (purple) and for datasets with the top 1/10/100/1000 binaries removed (different shades of green). We also show the total number of binaries in each bin ($N_{\rm tot}$, gray histogram). Vertical dashed lines indicate frequencies where the total number of binaries reach 1000 and 100. Note that the lines for the datasets without the top 1000/100 asymptote to the corresponding vertical lines, indicating frequency bins where we remove a large fraction of the binaries.}
 \label{fig:eff_num_binary}
\end{figure}

To quantify how much $h_c^{\rm GWB}$ is dominated by a few sources at a given frequency, we look at the minimum number of binaries contributing at least 90\% of $h_c^{\rm GWB}$ at a given frequency ($N_{90}$).
Figure \ref{fig:eff_num_binary} shows the median $N_{90}$ value over 20 realizations. We show this for the total population and also for populations where we removed the top $M=1/10/100/1000$ binaries. The total number of binaries in each frequency bin is also indicated. While the bulk of the GWB signal is built up from $\mathcal{O}(10^3)$ binaries at the lowest frequency bins, it is dominated by less than 10 binaries above about 100 nHz. This is expected from our qualitative assessment of the spectra in Fig.~\ref{fig:spectrum_variance}, where we have seen an increased scatter at high frequencies. We can also see that as we remove the brightest binaries from the population, the effective number of binaries contributing to the GWB increases. Note that as we remove more and more binaries, we can reach a limit where we are removing a significant fraction of the total number of binaries, and thus removing more binaries actually decreases $N_{90}$. See e.g.~``-1000'' and ``-100'' lines around 40 nHz in Fig.~\ref{fig:eff_num_binary}). We also show the frequencies where the total number of binaries reach 100 and 1000 (vertical dashed lines). Note that the ``-1000'' and ``-100'' lines approach these horizontal lines, indicating the point where we remove all the binaries in the given bin.

Figure \ref{fig:eff_num_binary} illustrates why modeling a realistic background as a GWB and a few individual sources is well motivated. Modeling all the binaries as a GWB is not justified, since the background level for the full population is dominated by a few sources at high frequencies, so the central limit theorem breaks down. However, if we only consider the population with hundreds of the brightest binaries removed, the background level is determined by a large ($\gtrsim 100-1000$) number of binaries, so treating it as a Gaussian background is a good approximation. In this case the non-Gaussian nature of the background is modeled by adding the brightest binaries as individual sources to the full signal (see Eq.~(\ref{eq:split_method})).

\subsection{\label{ssec:isotropy}Statistical isotropy}

\begin{figure*}[htbp]
   \centering
\begin{tabular}{cc}
\includegraphics[width=\columnwidth]{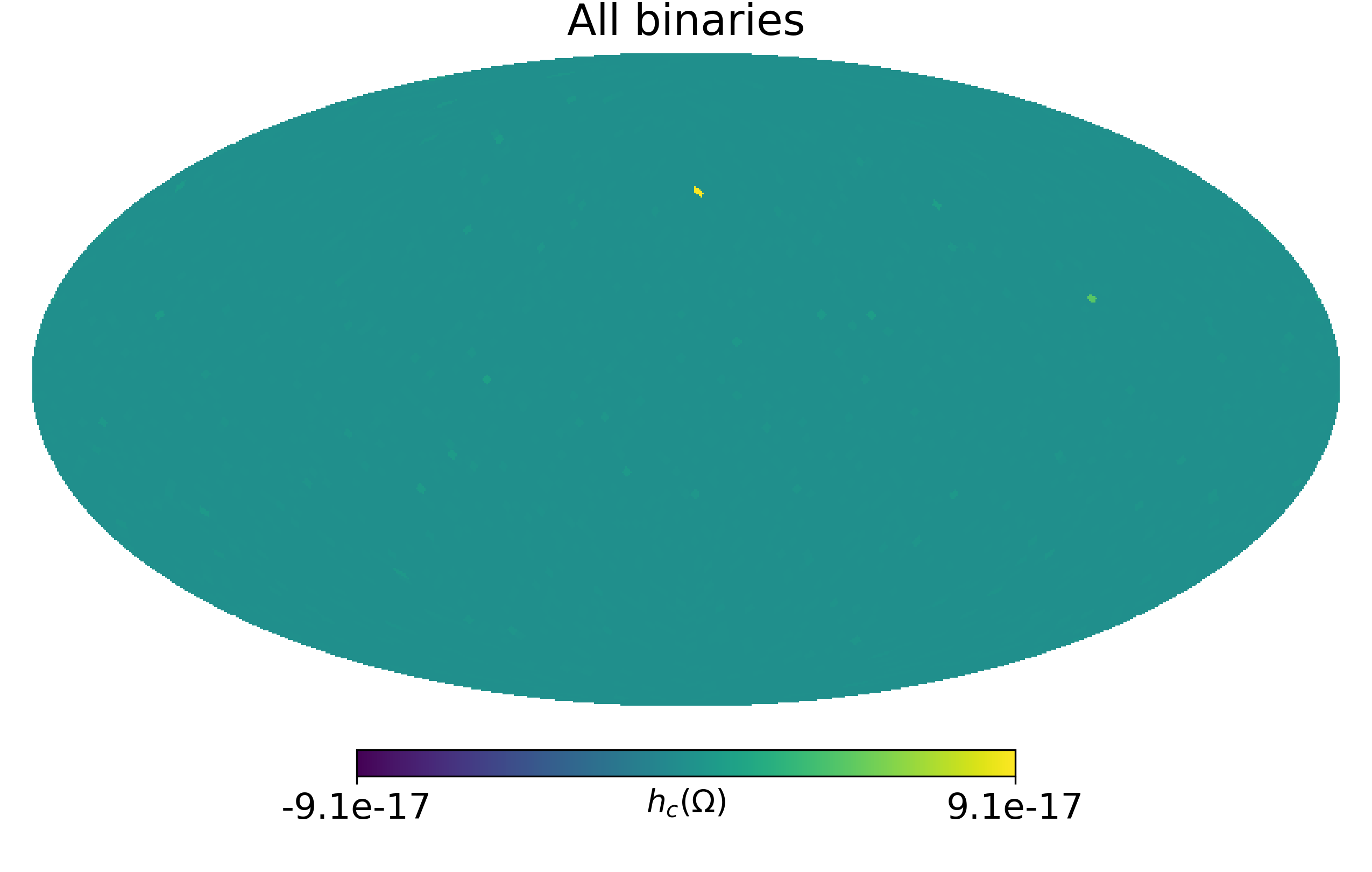}&
\includegraphics[width=\columnwidth]{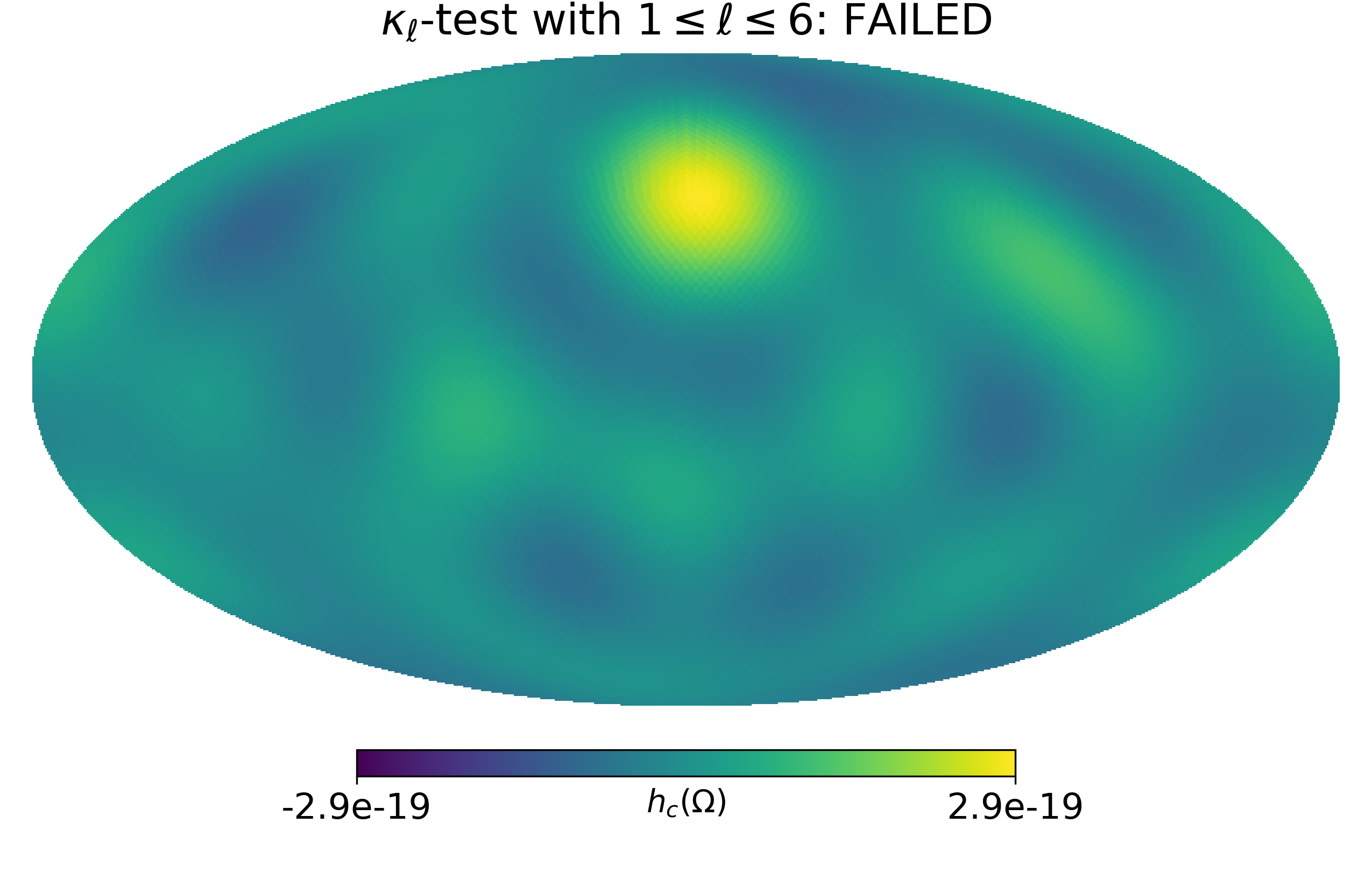}\\
\includegraphics[width=\columnwidth]{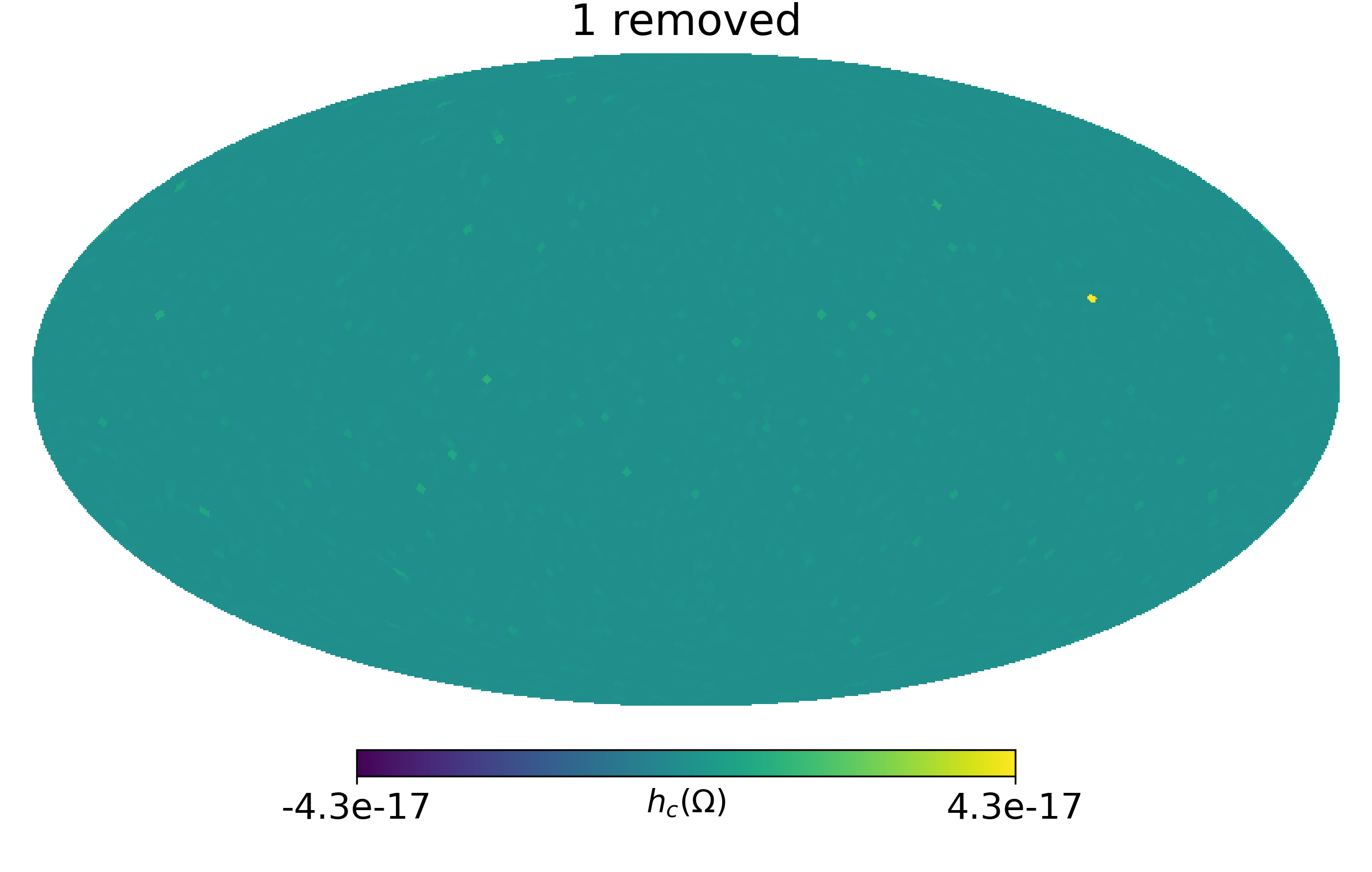}&
\includegraphics[width=\columnwidth]{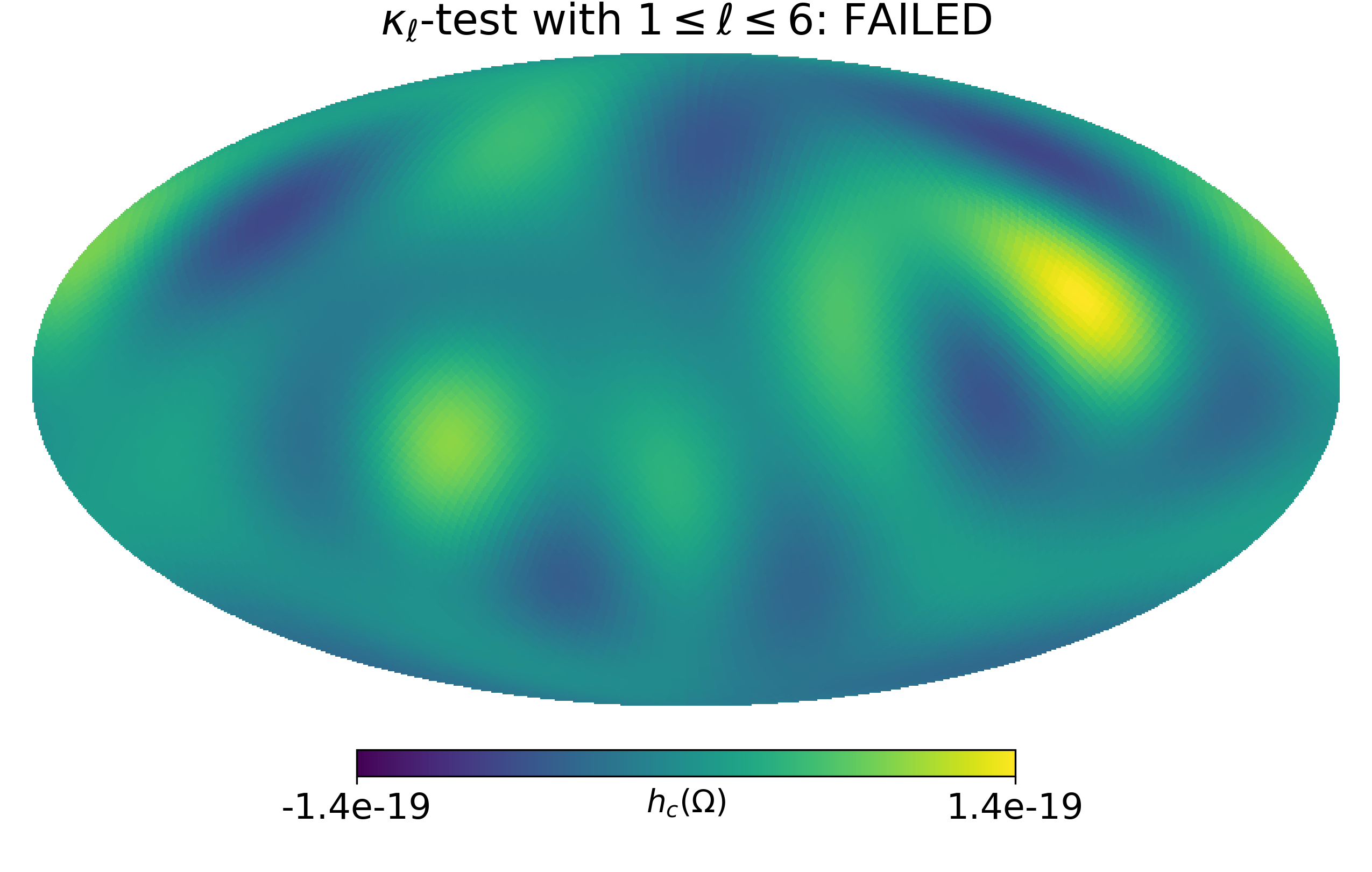}\\
\includegraphics[width=\columnwidth]{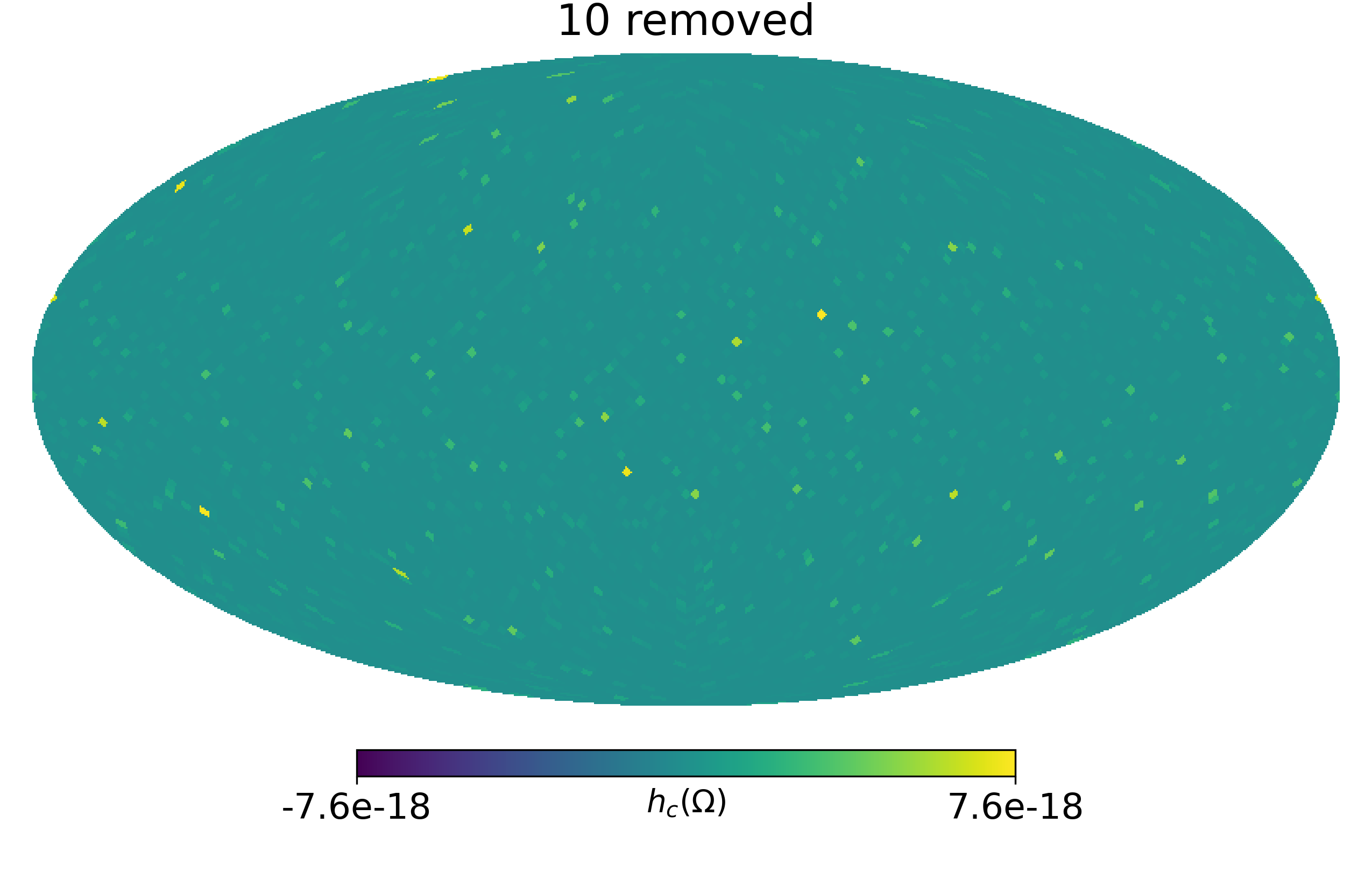}&
\includegraphics[width=\columnwidth]{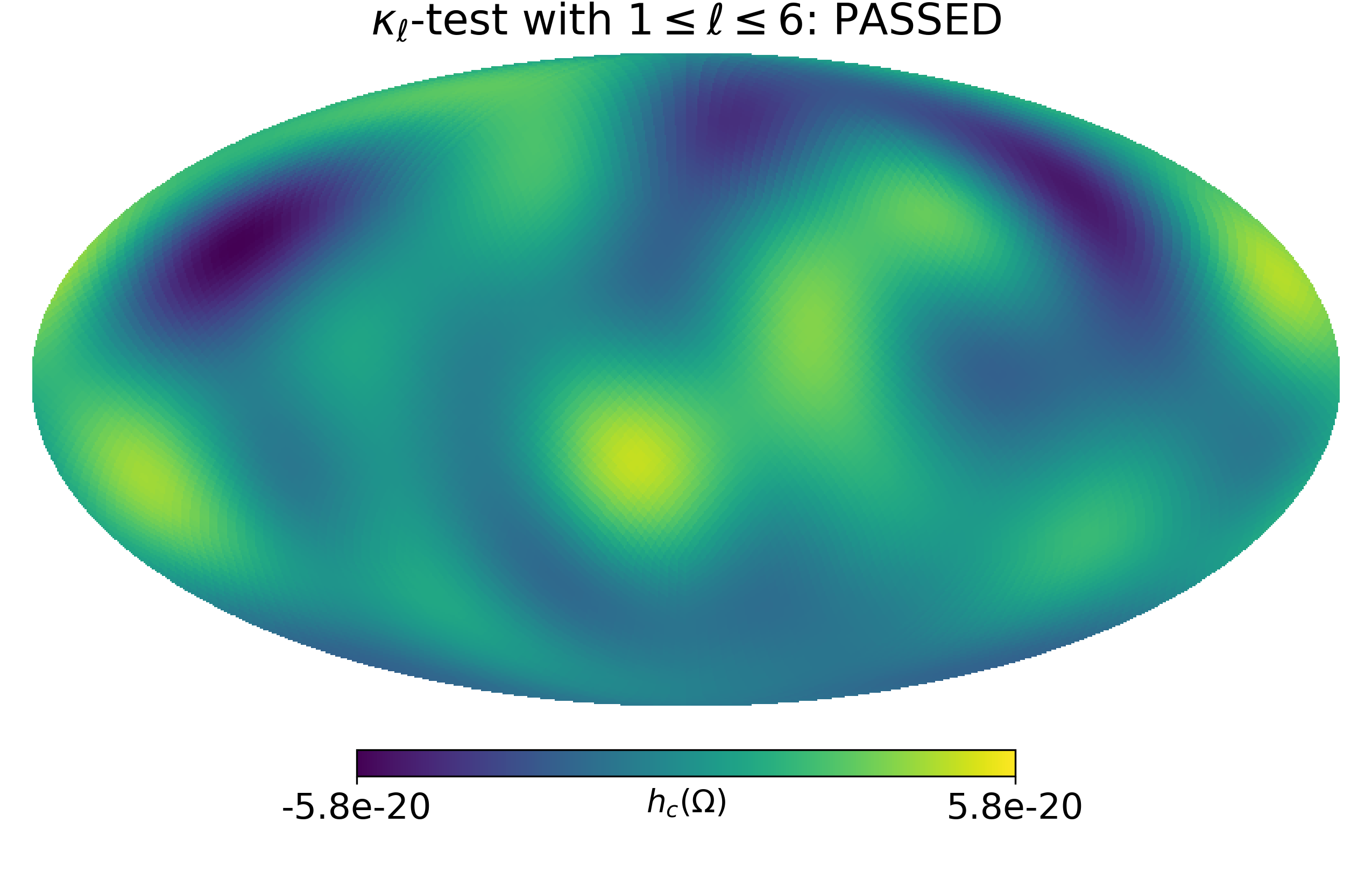}\\
\includegraphics[width=\columnwidth]{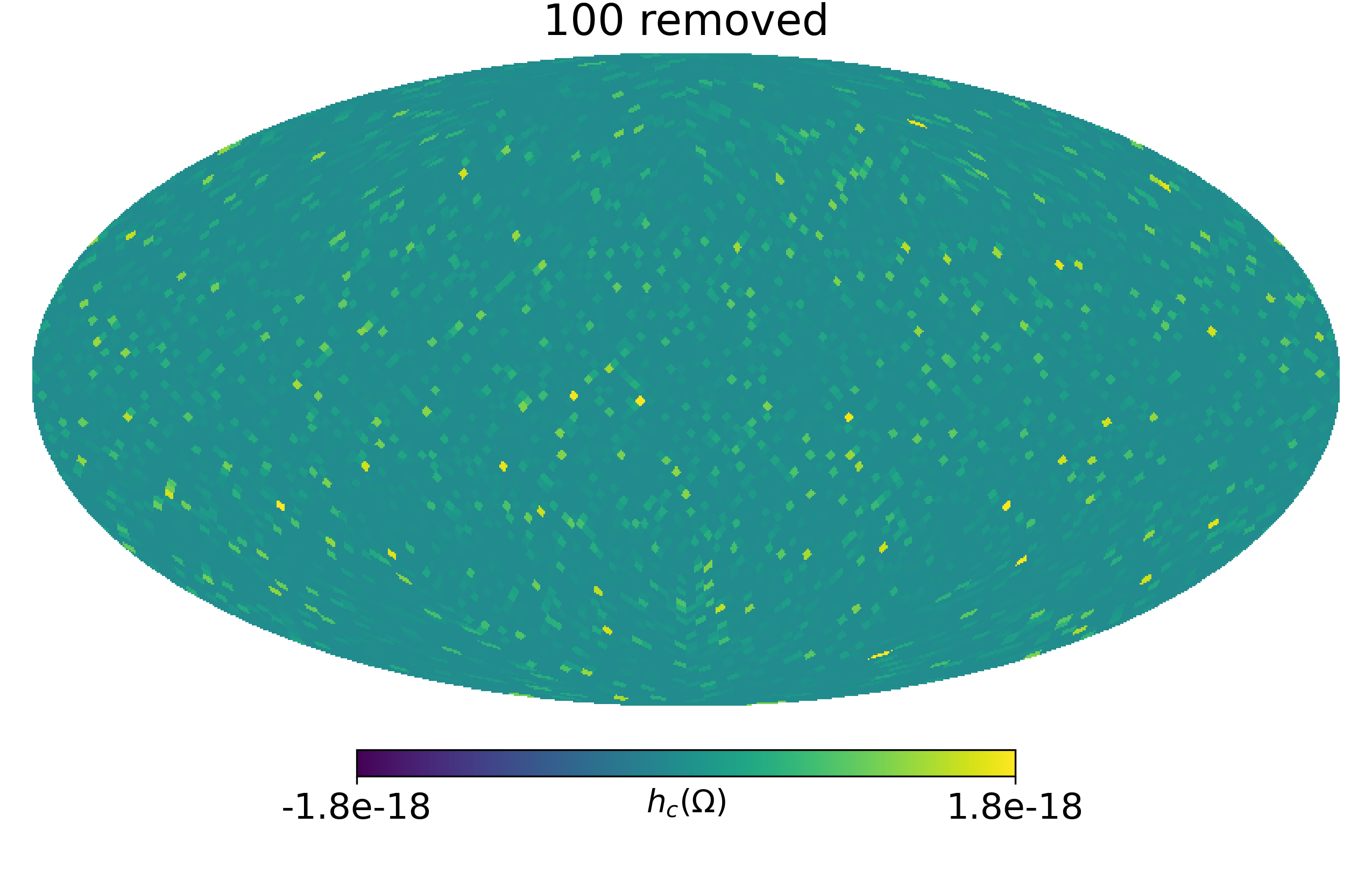}&
\includegraphics[width=\columnwidth]{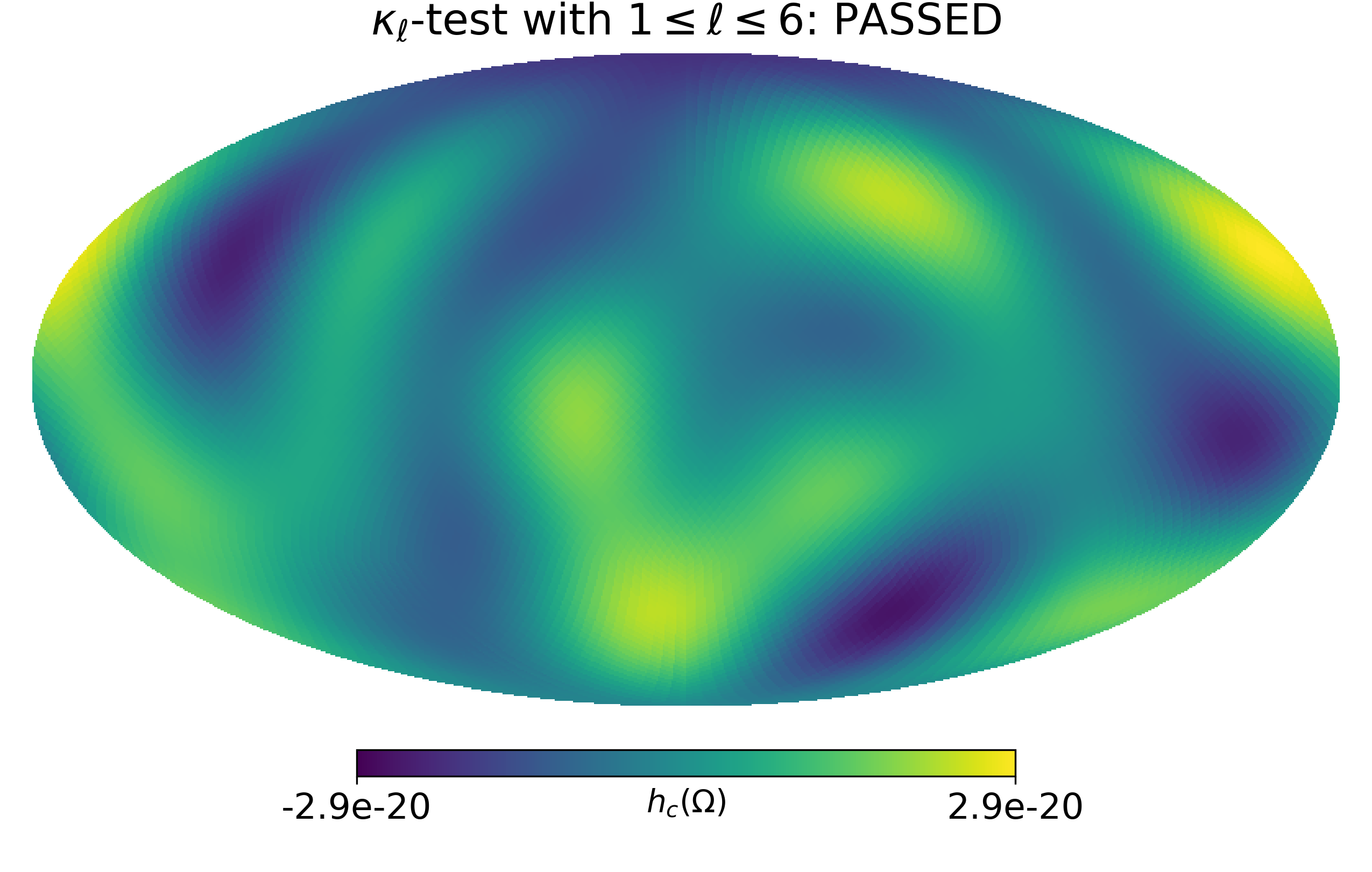}\\
\end{tabular}
    \caption{Angular distribution of the characteristic amplitude in the frequency bin with 66.5 nHz$<f_{\rm obs}<$68.7 nHz. We show sky maps for all binaries (first row), and with 1/10/100 brightest binaries removed (2nd/3rd/4th rows). The left column shows the raw maps, while the right column shows the maps reconstructed with the same range of angular scales as was used for the anisotropy test ($1\leq\ell\leq6$). Only the maps with 10 and 100 binaries removed passed the isotropy test in this example. The monopole term was subtracted from each map. Note the changing scale of the colormaps, which we forced to be symmetric around 0.}
    \label{fig:skymaps} 
\end{figure*}

In the simplest description, it is also usually assumed that the GWB exhibits statistical isotropy. However, this is not expected to be true for realistic backgrounds. SMBHBs are hosted by galaxies, which cluster into galaxy clusters, so nearby galaxy clusters can result in an overabundance of nHz GW sources in a particular sky location. In addition, SMBHBs can produce anisotropy even when they are uniformly distributed in volume, if a small number of them dominate the nHz GW sky. This is in stark contrast with a GWB of primordial cosmological origin, which would be perfectly statistically isotropic, so the origin of the GWB can be firmly established by detecting anisotropy \citep{Chiara_anisotropy_2013}. Indeed, there are numerous methods to search for signatures of anisotropy in the GWB \citep{Neil_anisotropic_search, Chiara_anisotropy_2013, Steve_anisotropy_2013, BipoSH_anisotropy, Steve_bumpy_background, Chiara_anisotropy_2020}, and by some estimates PTAs might be able to detect anisotropy just a few years after detecting the isotropic component of the GWB \citep{Nihan_anisotropy_forecast}.

In our modeling we assume a uniform in volume distribution of sources. This means that the following results on anisotropy can be taken as a lower limit, since the effect of local inhomogeneities would further increase the level of anisotropy. To model the angular distribution of GW power, we pixelate the sky using the \texttt{HEALPix} framework \citep{healpix}. Since this produces equal-area pixels, randomly assigning each SMBHB to a pixel is equivalent to distributing them uniformly on the sky. We carry out our analysis independently for each frequency bin. For each pixel, we sum up $h_c^2$ values of all the binaries assigned to that pixel, resulting in a discretized function of sky location $h_c^2 (\Omega)$.

Figure \ref{fig:skymaps} shows that distribution for a particular realization at the frequency bin between 66.5 nHz and 68.7 nHz. This frequency bin has a total number of 9,801 binaries. The different rows show maps with different numbers of the brightest binaries removed. For each of these we show both the raw map of $h_c (\Omega)$ (left) and a reconstructed map with angular scales restricted to those used for the anisotropy test described below (right). Note that the mean amplitude has been subtracted to better show the fluctuations. We allow the colorscales to be different for each sky map, but force it to be symmetric around 0. This particular example shows a highly anisotropic distribution due to a few brightest binaries dominating the sky, but as more and more of the binaries are excised, the remaining GWB gets more and more isotropic. By the time we remove the top 10 binaries, the GWB sky map restricted to large angular scales is visibly indistinguishable from an isotropic distribution. This is similar to the findings of previous studies reported in \citet{Steve_anisotropy_2013} and \citet{chiara_local_gw_landscape}, where authors found that small angular scale power can be reduced (and thus the map can be made more isotropic) by removing the brightest source (see Fig.~2 in both references).

To quantitatively assess the statistical isotropy of maps like those shown in Fig.~\ref{fig:skymaps}, we follow the formalism introduced in \citet{kappa_test_2003} and \citet{kappa_test_formalism}. This method was used to test the statistical isotropy of the cosmic microwave background using data from the Wilkinson Microwave Anisotropy Probe \citep{kappa_test_Neil}. We start by decomposing our maps in terms of $Y_{\ell m}$ spherical harmonics:
\begin{equation}
h_c(\Omega) = h_c(\theta ,\varphi) = \sum _{\ell =0}^{\ell_{\rm max} }\sum _{m=-\ell }^{\ell } a_{\ell m}Y_{\ell m}(\theta ,\varphi ).
\end{equation}
The $a_{\ell m}$ spherical harmonic coefficients can be used to compute the bipolar spherical harmonic coefficients:
\begin{equation}
A_{ll'}^{\ell M} = \sum _{mm'} a_{lm} a_{l'm'} \mathfrak{C}_{lml'm'}^{\ell M},
\end{equation}
where $\mathfrak{C}_{lml'm'}^{\ell M}$ are Clebsch-Gordan coefficients. The bipolar power spectrum is defined as:
\begin{equation}
\kappa_{\ell} = \sum _{ll'M} \left| A_{ll'}^{\ell M} \right|^2.
\end{equation}
Note that for statistically isotropic maps the usual angular power spectrum $C_l=1/(2l+1) \sum_m |a_{l m}|^2$ contains all information about the map, and we have $A_{ll'}^{\ell M} = (-1)^l C_l \sqrt{2l+1} \delta_{ll'} \delta_{\ell 0} \delta_{M 0}$ and $\kappa_{\ell} = \kappa_0 \delta_{\ell 0}$. After correcting for the estimation bias (Eq.~(14) of \citet{kappa_test_2003}), $\kappa_{\ell}$ has an expectation value of 0 for all $\ell\geq1$ under statistical isotropy. The variance can also be analytically computed (Eq.~(17) of \citealt{kappa_test_2003}), so one can test for statistical isotropy by checking if $\kappa_{\ell}$ are consistent with zero within their theoretical variance.

We calculate $\kappa_{\ell}$ for $1\leq\ell\leq6$, which corresponds to probing angular scales larger than 30$^{\circ}$. The cut-off at $\ell=6$ is motivated by the angular resolution sensitivity forecasts from \citet{Nihan_anisotropy_forecast}.
We apply a Gaussian low-pass filter on the $a_{\ell }^{m}$ values as defined in Eq.~(5) of \citet{kappa_test_Neil} with $l_s=20$. This ensures that there is no spectral leakage affecting the analysis.
A key difference between our analysis and the one presented in \citet{kappa_test_Neil} is that we do not use the theoretical variance of $\kappa_{\ell}$ to decide if a map is isotropic or not. Instead, we compare the measured $\kappa_{\ell}$ values to $\kappa_{\ell}$ values calculated from $N_{\rm iso}$ realizations of a simulated isotropic map with the same angular power spectrum. We call a map statistically isotropic if the measured $\kappa_{\ell}$ values are within the maximum and minimum $\kappa_{\ell}$ from the truly isotropic maps for all $\ell$. This procedure allows us to set the false positive and false negative probabilities of our test by choosing the value of $N_{\rm iso}$. As discussed below, we perform multiple trials of this test, so we want to make sure that anisotropic maps rarely pass the test by chance. We choose $N_{\rm iso}=5$, which results in the false positive and false negative rates reported in Table \ref{tab:kappa_error_rates}. Note that increasing $N_{\rm iso}$ would further decrease the probability of anisotropic maps passing the test at the cost of fewer truly isotropic maps passing it.

\begin{table}[htbp]
\centering
\caption{\label{tab:kappa_error_rates}%
False negative and false positive probabilities of the anisotropy test with $N_{\rm iso}=5$.
}
\begin{tabular}{c|cc}
Isotropy test result & Pass & Fail\\
\hline
Truly isotropic map & 57\% & 43\% \\
Truly anisotropic map\footnote{Based on randomly selected map dominated by a single source.} & 12\% & 88\% \\
\end{tabular}
\end{table}

We performed $\kappa_{\ell}$-tests on all four maps shown on the right column of Figure \ref{fig:skymaps}. The maps with all binaries and with the single brightest binaries removed failed the test, and are thus not consistent with being statistically isotropic. The maps with 10 and 100 binaries removed passed the test. It is instructive to compare the maps in the left and right columns. Based on the raw maps we might say that even the map with 10 binaries removed is not isotropic. However, the maps on the right correspond to what the $\kappa_{\ell}$-test actually sees. It is evident even in those large angular scale features that the first two maps are dominated by a few sources. Note in particular that these maps are highly asymmetric around their mean values resulting in the lack of dark blue colors. However, once we remove the 10 brightest binaries, the map is consistent with statistical isotropy up to $\ell=6$. This example reinforces our previous qualitative assessment that removing the few brightest binaries makes the GWB more isotropic.

\begin{figure}[htb]
 \centering
   \includegraphics[width=\columnwidth]{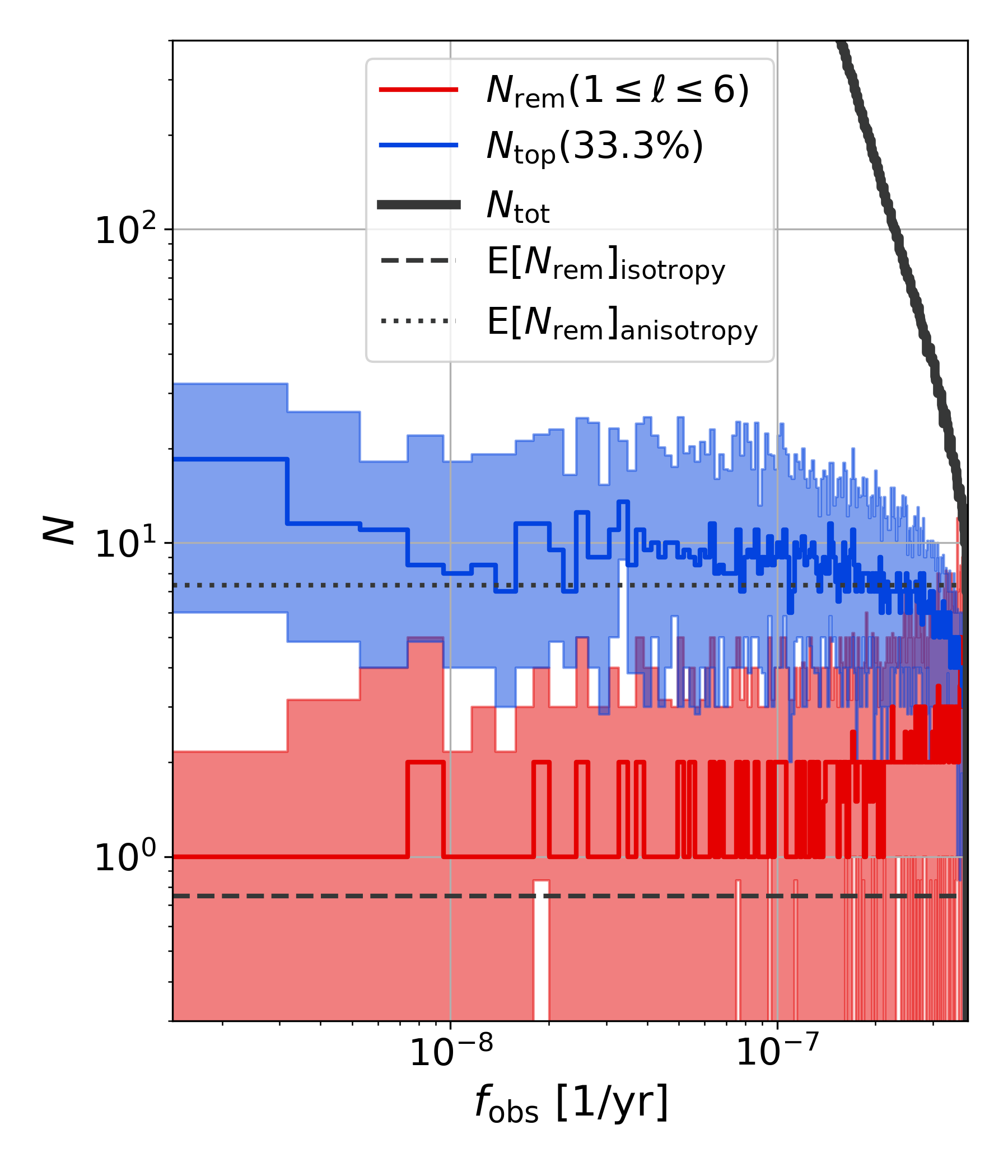}
 \caption{Minimum number of sources removed to achieve statistical isotropy based on the $\kappa_{\ell}$-test with $\ell$ between 1 and 6, $N_{\rm rem} (1 \leq \ell \leq 6)$. We show the median and 68\% credible interval of $N_{\rm rem}$ over 30 GWB realizations. We also show the number of binaries with an $h_c$ within a factor of 3 of the brightest binary in the frequency bin, $N_{\rm top} (33.3\%)$. Note how statistical isotropy is reached when $N_{\rm top} (33.3\%) \sim 10$. The black line shows the total number of sources in each frequency bin. The dashed (dotted) horizontal line shows the expected value of $N_{\rm rem}$ due to the false negative (positive) probability of our test (see Table \ref{tab:kappa_error_rates}) if all maps are isotropic (anisotropic).}
 \label{fig:KS_test_min_remove}
\end{figure}

We can ask how many binaries one needs to remove to get a background consistent with statistical isotropy. To answer that we start with a $\kappa_{\ell}$-test on a map with all binaries, and remove binaries one at a time until we reach a statistically isotropic map. We call the number of binaries removed to achieve this $N_{\rm rem}$. We repeat this procedure over all frequency bins and multiple realizations (see Fig.~\ref{fig:KS_test_min_remove}). We can see that there is a trend of needing to remove more binaries as we move to higher frequencies. Figure~\ref{fig:KS_test_min_remove} also shows the number of binaries with an amplitude within a factor of three of the brightest binary, $N_{\rm top} (33.3\%)$, after removing the $N_{\rm rem}$ brightest binaries. We can see that we typically reach isotropy when there are more than $\sim 10$ binaries with comparable amplitudes. This is consistent with the fact that at $\ell=6$ there are 13 different spherical harmonics, so once there are $\gtrsim10$ comparable sources randomly placed on the sky, we cannot distinguish that from an isotropic sky. Note that at the highest frequencies $N_{\rm rem}\simeq N_{\rm tot}$, where $N_{\rm tot}$ is the total number of binaries in each frequency bin. This is because in our procedure we set $N_{\rm rem}=N_{\rm tot}$ if we cannot achieve statistical isotropy with any value of $N_{\rm rem}$. This is a similar feature to the one in Fig.~\ref{fig:eff_num_binary}, where removing more binaries does not get us closer to a true stochastic background when $N_{\rm tot}$ is small to start with. We also show the expected value of $N_{\rm rem}$ assuming all maps are isotropic, $\mathrm{E}[N_{\rm rem}]_{\rm isotropy}=\alpha/(1-\alpha)=0.75$, which is based on $N_{\rm rem}$ following a negative binomial distribution with the success probability of $\alpha=0.43$ corresponding to the false negative probability of the $\kappa_{\ell}$-test reported in Table \ref{tab:kappa_error_rates}. We can see that at the lowest frequencies, the median value of $N_{\rm rem}$ is only slightly above this expectation value, suggesting that a fraction of these maps are isotropic even without removing any bright binaries. We also show the expected value of $N_{\rm rem}$ assuming every map is anisotropic, $\mathrm{E}[N_{\rm rem}]_{\rm anisotropy}=7.33$, which is higher than the median $N_{\rm rem}$ at any frequency, and thus not expected to significantly affect the results.

\subsection{\label{ssec:hd-variance}Variance of the Hellings-Downs curve}

It can be shown that an isotropic, unpolarized, time-stationary, Gaussian GWB produces residuals that are correlated between pulsar pairs, and the amount of correlation between pulsar $i$ and pulsar $j$ is described by the HD curve \citep{HD}\footnote{Note that we follow the normalization convention where $\Gamma_{ii}=1$.}:
\begin{equation}
 \Gamma_{ij} = \frac{1}{2} - \frac{1}{4} x_{ij} + \frac{3}{2} x_{ij} \ln (x_{ij}) + \frac{1}{2} \delta_{ij},
\label{eq:HD}
\end{equation}
where the term with $\delta_{ij}$ encodes correlations due to the pulsar terms and it vanishes when looking at different pulsars, and $x_{ij}$ is defined as:
\begin{equation}
 x_{ij} = \frac{1-\cos \gamma_{ij}}{2},
\end{equation}
where $\gamma_{ij}$ is the angular separation between pulsars $i$ and $j$. It has been shown in \citet{Neil_CW_HD} that Eq.~(\ref{eq:HD}) not only holds for GWBs, but also for a so called continuous-wave (CW) signal with a quasi-constant frequency and amplitude from an individual SMBHB.

While Eq.~(\ref{eq:HD}) predicts the expectation value of the correlations as a function of $\gamma$, individual correlations can have a significant variance (see \citet{BruceAllen_HD_variance} and references therein). It has been suggested that the way the variance varies with angle might be used to distinguish between primordial and astrophysical sources of the GWB \citep{BruceAllen_HD_variance}.

There are two sources of that variance that are present even in idealized noise-free observations, which we can call pulsar variance and cosmic variance, following the nomenclature of \citet{BruceAllen_HD_variance}. Pulsar variance arises from the fact that individual correlations can depend not only on $\gamma$, but also on three other angles describing the position of the two pulsars on the sky. Having a large number of pulsar pairs, one can in principle average out the pulsar variance. Cosmic variance refers to statistical deviations from Eq.~(\ref{eq:HD}) in a given realization of the GWB. Unlike pulsar variance, we cannot average out cosmic variance, since we only have one universe to make observations of.

To quantify the variance of the HD correlations, we performed three different simulations, each with 100 fully independent realizations:
\begin{enumerate}[label=(\roman*)]
    \item GWB: an isotropic stationary Gaussian stochastic background with a characteristic strain amplitude of $1\times 10^{-15}$ measured at $f_{\rm obs}=1 \ {\rm year}^{-1}$ and spectral index of -2/3;
    \item CW: GW signal of an individual SMBHB with fixed amplitude, $\mathcal{M}=10^9 \ M_{\odot}$, $f_{\rm obs}=10$ nHz, and nuisance parameters drawn randomly from uniform distributions;
    \item Population: realistic GWB simulated as in Eq.~(\ref{eq:split_method}).
\end{enumerate}
For these simulations we use the sky location of the 67 pulsars in the NANOGrav 15-year dataset \citep{nanograv_15_data}, and simulate evenly sampled observations over 15 years with a 30 day cadence and with no noise. From the resulting $r_k(t_i)$ residuals at times $t_i$ in pulsar $k$ we calculate the zero-lag correlation between each pulsar pair as:
\begin{equation}
\mathcal{C}_{kl} = \frac{1}{N_{\rm obs}} \sum_{i=1}^{N_{\rm obs}} r_k(t_i) r_l(t_i),
\end{equation}
where $N_{\rm obs}$ is the number of observations per pulsar. We also introduce:
\begin{equation}
\hat{\mathcal{C}}_{kl} = \mathcal{C}_{kl} \left( \frac{1}{N_{\rm PSR}} \sum_{k=1}^{N_{\rm PSR}} \mathcal{C}_{kk} \right)^{-1},
\end{equation}
which normalizes $\mathcal{C}_{kl}$ with the auto-correlation terms, so that it follows the same convention as Eq.~(\ref{eq:HD}).

\begin{figure}[htb]
 \centering
   \includegraphics[width=\columnwidth]{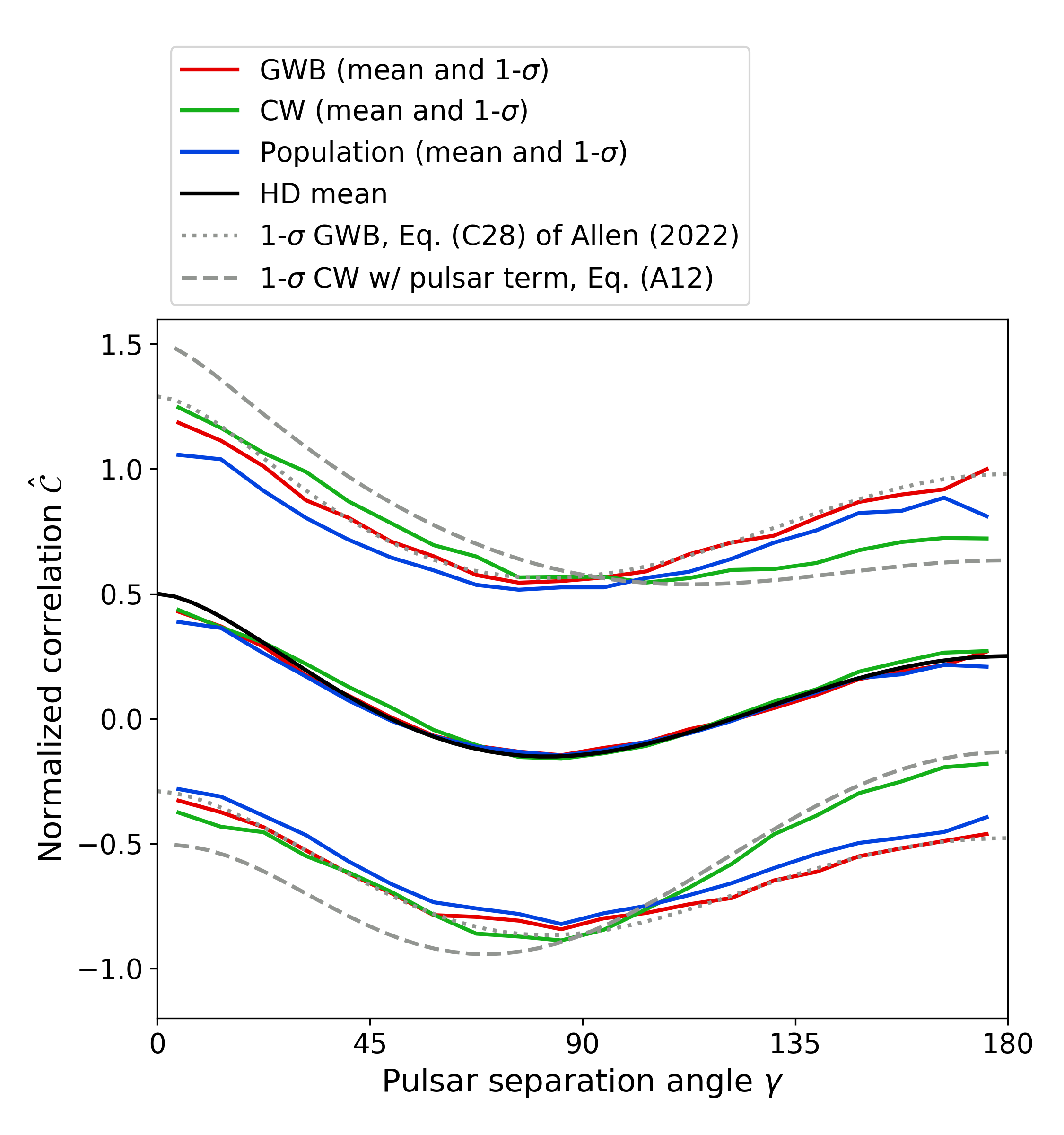}
 \caption{Mean and variance of inter-pulsar correlations. We show results for an idealized GWB, a single CW source, and a realistic population-based GWB. We also show the theoretically expected mean and variance of the correlations under different assumptions. The small discrepancy between the theoretical and simulated CW curves is due to the anisotropic distribution of pulsars used for the simulation (see Appendix \ref{sec:hd_variance_appendix}). Note that we show the results in 20 angular bins, and the results were averaged over 100 realizations to reduce the scatter of these curves.}
 \label{fig:HD-variance-unbinned}
\end{figure}

Figure \ref{fig:HD-variance-unbinned} shows the mean and 1-$\sigma$ region of $\hat{\mathcal{C}}_{kl}$ as a function of $\gamma$. The means follow the expected HD curve for all three simulations\footnote{This is expected since we are showing the mean over many realizations, which removes the cosmic variance from the mean curves. For a given realization, we expect the mean to be different from the HD curve by an amount predicted by the cosmic variance.}. We also show the theoretical variance for a purely stochastic background from \citet{BruceAllen_HD_variance}, and for a single CW with pulsar terms from Appendix \ref{sec:hd_variance_appendix} based on results from \citet{BruceAllen_HD_variance}. We can see that our results are in good agreement with these predictions. In particular, the GWB curve follows the prediction for an isotropic Gaussian background perfectly. The Population simulation also follows the shape of the same theoretical curve, but with a slightly lower variance at all angles. This constant shift is most likely due to its different spectral properties, which influence the normalization between the mean and the variance of the correlations (i.e.~Eqs.~(C30) and (C32) of \citealt{BruceAllen_HD_variance}). This also seems to suggest that our realistic GWB behaves more like an idealized GWB than a single CW in this regard. The CW simulations also show some agreement with the corresponding theoretical predictions. In particular, these correctly predict that a single source has less variance than a GWB at large angles, but more at small angles. The deviation can be explained by the anisotropic distribution of pulsars on the sky, since they agree very well for a simulation with isotropically distributed pulsars (see Appendix \ref{sec:hd_variance_appendix}).

\begin{figure}[htb]
 \centering
   \includegraphics[width=\columnwidth]{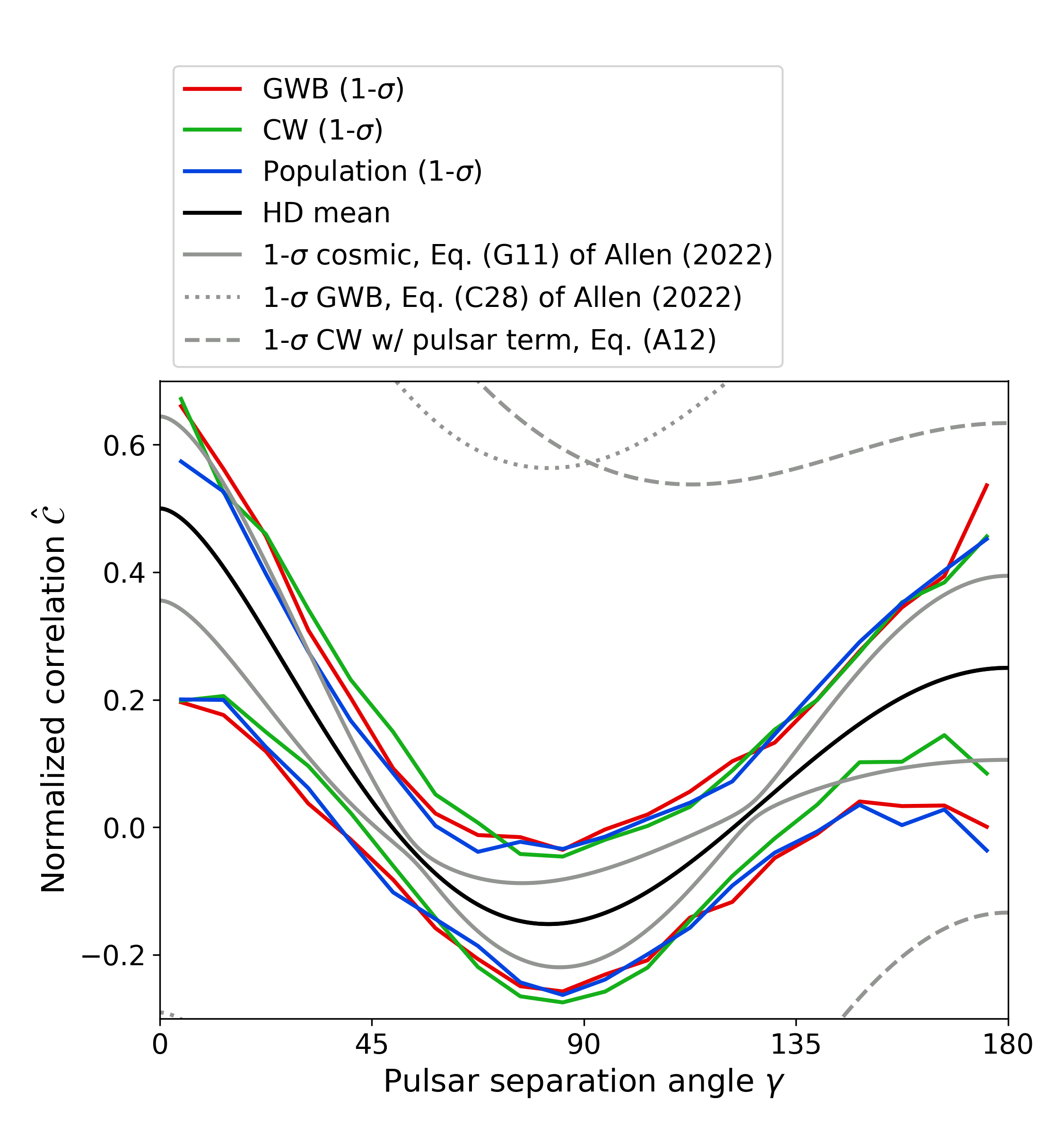}
 \caption{Variance of the mean correlations over 100 realizations. We show results for an idealized GWB, a single CW source, and a realistic population-based GWB. Besides the same theoretical curves as in Fig.~\ref{fig:HD-variance-unbinned}, we also show the theoretical variance from cosmic variance only \citep{BruceAllen_HD_variance}. Note the different vertical scale compared to Fig.~\ref{fig:HD-variance-unbinned}.}
 \label{fig:HD-variance-binned}
\end{figure}

The 1-$\sigma$ bounds we see in Fig.~\ref{fig:HD-variance-unbinned} include both pulsar and cosmic variance. To reduce the pulsar variance one can take the mean correlation in a set of angular separation bins. In the limit of large number of pulsars, such an averaging should remove the pulsar variance, so the only source of variance is the cosmic variance (assuming noiseless observations). To test this scenario we analyzed the same simulations as above, but instead of calculating the variance of individual correlations, we first took the mean of the correlations in each realization, and then calculated the variance of those means over all realizations. The results are shown in Figure \ref{fig:HD-variance-binned}, where we used 20 equal-sized angular bins for the averaging. We also show the same theoretically expected variances as in Fig.~\ref{fig:HD-variance-unbinned}. In addition, solid gray lines indicate the cosmic variance (Eq.~(G11) in \citealt{BruceAllen_HD_variance}). As expected, the variance of the correlations is drastically reduced by the binning and averaging procedure. However, the variance is still larger than the cosmic variance at some angular separations. This is due to the finite number of pulsar pairs available for averaging in our simulations. Increasing the bin sizes used for the averaging brings these lines closer to the theoretical curves at the cost of losing angular resolution. Also note that the results from our three simulations are all consistent with each other within their statistical uncertainty in this averaged representation. In this exercise we weighted all pulsar pairs equally in the averaging. \citet{allen_romano_cosmic_variance} derived the optimal way of reducing the pulsar variance, which does not use equal weights for different pulsar pairs.

\begin{figure}[htb]
 \centering
   \includegraphics[width=0.9\columnwidth]{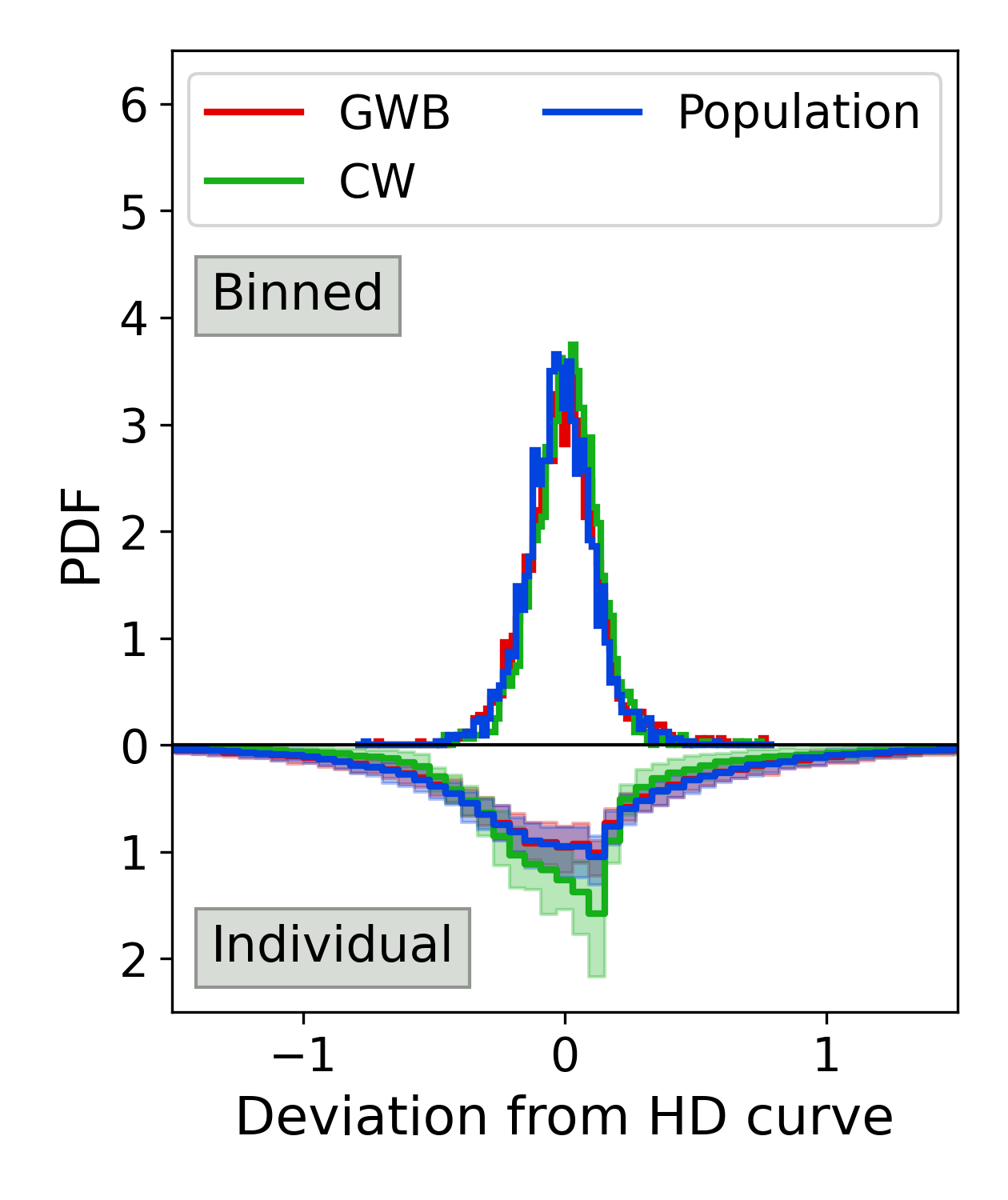}
 \caption{Histogram of deviations from the HD curve for all three simulations we performed. The bottom panel shows the mean and 90\% credible interval for the distribution of all individual correlation values. The top panel shows the distribution of mean correlation values calculated in each angular bin and each realization.}
 \label{fig:HD-diff-hist}
\end{figure}

Figures \ref{fig:HD-variance-unbinned} and \ref{fig:HD-variance-binned} only show the first and second moments of the distributions of correlations. However, one advantage of our direct simulation approach over analytical calculations is that we can easily produce the full distribution of correlations. To that end, we show the distribution of the deviation from the HD curve both with and without binning in Figure \ref{fig:HD-diff-hist}. We can see that the binned correlations show a consistent distribution for all three of our simulations. This is not surprising given that their means and variances are indistinguishable, as we have seen in Fig.~\ref{fig:HD-variance-binned}. As expected, the individual correlations show a significantly wider distribution. We can also see that while the population-based and idealized GWB simulations show almost identical distributions, the histogram is significantly different for the simulations with an individual CW signal. As it was previously pointed out in \citet{Neil_CW_HD}, the overall distribution of correlation deviations has a heavier tail for a stochastic background then for a single CW source. Also note that the distribution of individual correlations shows significant non-Gaussian features, as it is expected based on analytical calculations (see footnote [40] in \citealt{BruceAllen_HD_variance}).

\section{\label{sec:detectability}Prospects for distinguishing individual binaries}

Our method of modeling realistic backgrounds is also uniquely suitable to study detection prospects of individual binaries in the presence of a stochastic background. By setting $M=1$ in Eqs.~(\ref{eq:split_method}) and (\ref{eq:gwb_sqsum}) we can separate the single brightest binary and the rest of the GWB in each bin. Thus we can take into account the confusion noise coming from the GWB itself, along with the white and red noise in each pulsar. Then we can calculate the expected signal-to-noise ratios (SNRs) for those outlier sources as (see e.g.~Eq.~(219) of \citealt{Neil_book}):
\begin{equation}
{\rm SNR} = (s(t; \boldsymbol\theta)|s(t; \boldsymbol\theta))^{1/2},
\end{equation}
where $s(t; \boldsymbol\theta)$ is the CW signal in question, and we define the inner product on residuals as:
\begin{equation}
(a|b) = a^T C^{-1} b,
\end{equation}
where $C$ is the noise covariance matrix that takes into account white and red noise in each pulsar and a stochastic background based on Eq.~(\ref{eq:gwb_sqsum}) with $M=1$. We calculate inner products using the \texttt{ENTERPRISE} software package \citep{enterprise}.

\subsection{\label{ssec:snr_variance}Variance of SNR over extrinsic parameters}
\begin{figure}[htb]
 \centering
   \includegraphics[width=0.9\columnwidth]{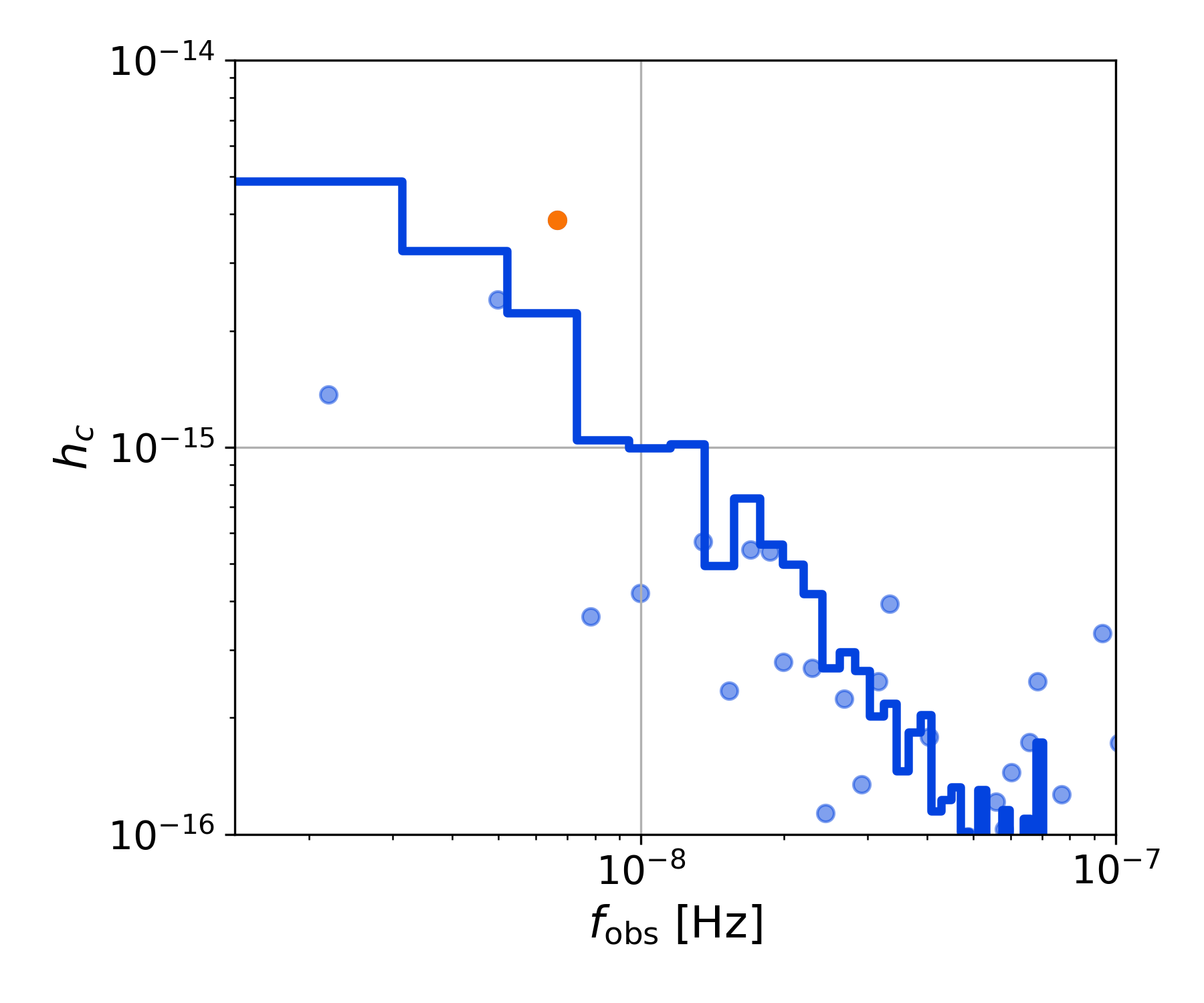}
 \caption{GWB spectrum in a particular realization (blue histogram) and the brightest binary in each bin (blue dots). The binary we focus our attention on is indicated by the orange dot.}
 \label{fig:spectrum_example}
\end{figure}

The SNR can have a significant variance even when we fix the intrinsic parameters of the source. These are due to effects of extrinsic parameters: inclination angle, polarization angle, initial phase, and two angles describing the sky location of the source. To explore the effect of these on SNR, we focus on a particular binary with fixed amplitude/distance, chirp mass, and GW frequency, and calculate the distribution of SNRs over extrinsic parameters. Figure \ref{fig:spectrum_example} shows this binary on the $h_c$--$f_{\rm obs}$ plane (orange dot). We also show the GWB spectrum for this particular realization (blue histogram), and the single brightest binary in each frequency bin (blue dots). Note that we chose our binary to be the one that produces the highest SNRs in this realization. We use a simulated dataset based on the properties of pulsars in the NANOGrav 12.5-year dataset \citep{nanograv_12p5_data}, and we extend the observation time to 15 years, as it was done in \citet{Astro4Cast}.

\begin{figure}[htb]
 \centering
   \includegraphics[width=\columnwidth]{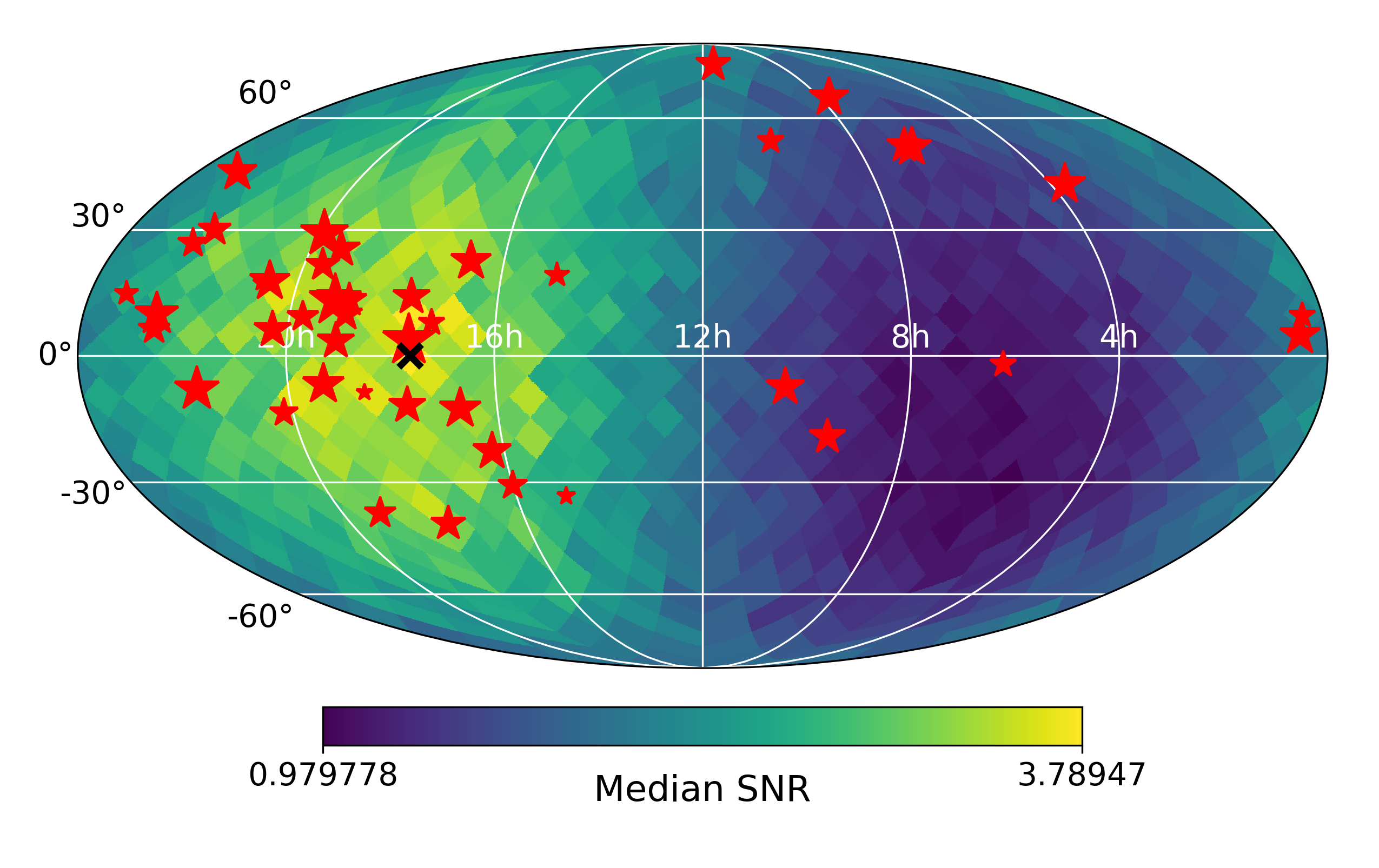}
 \caption{Distribution of the median SNR over the sky for a particular source. Red stars indicate the locations of pulsars in our array (based on the NANOGrav 12.5-year dataset, see \citealt{nanograv_12p5_data}), with their sizes corresponding to their sensitivity as defined in Eq.~(\ref{eq:sensitivity}). The black cross indicates the sky pixel with the highest median SNR.}
 \label{fig:snr_sky}
\end{figure}

Figure \ref{fig:snr_sky} shows the distribution of the median SNR over the sky for this particular source. We also show the location of the pulsars in our simulated array as red stars. The sizes of the stars are scaled by a rough estimate of the sensitivity of the pulsars defined as:
\begin{equation}
{\rm Sensitivity} \propto \left( \sum_{i=1}^{N_{\rm obs}} \frac{1}{(\Delta t_i)^2}  \right)^{1/2},
\label{eq:sensitivity}
\end{equation}
where $\Delta t_i$ is the nominal TOA error of the $i$th observation. Note that this simple expression reproduces the expected scaling both with the number of observations ($\sim \sqrt{N_{\rm obs}}$) and with the TOA errors ($\sim 1/\Delta t_i$). We can see in Fig.~\ref{fig:snr_sky} that the sensitivity on the sky shows a dipolar structure, where one gets SNRs almost a factor of four higher in one direction than the antipodal direction. This is in agreement with the highly anisotropic upper limits found by the search for individual binaries in the NANOGrav 11-year dataset \citep{nanograv_11_cw}. The reason for this is the highly anisotropic distribution of NANOGrav pulsars, which tend to be concentrated around the galactic center.

\begin{figure}[htb]
 \centering
   \includegraphics[width=\columnwidth]{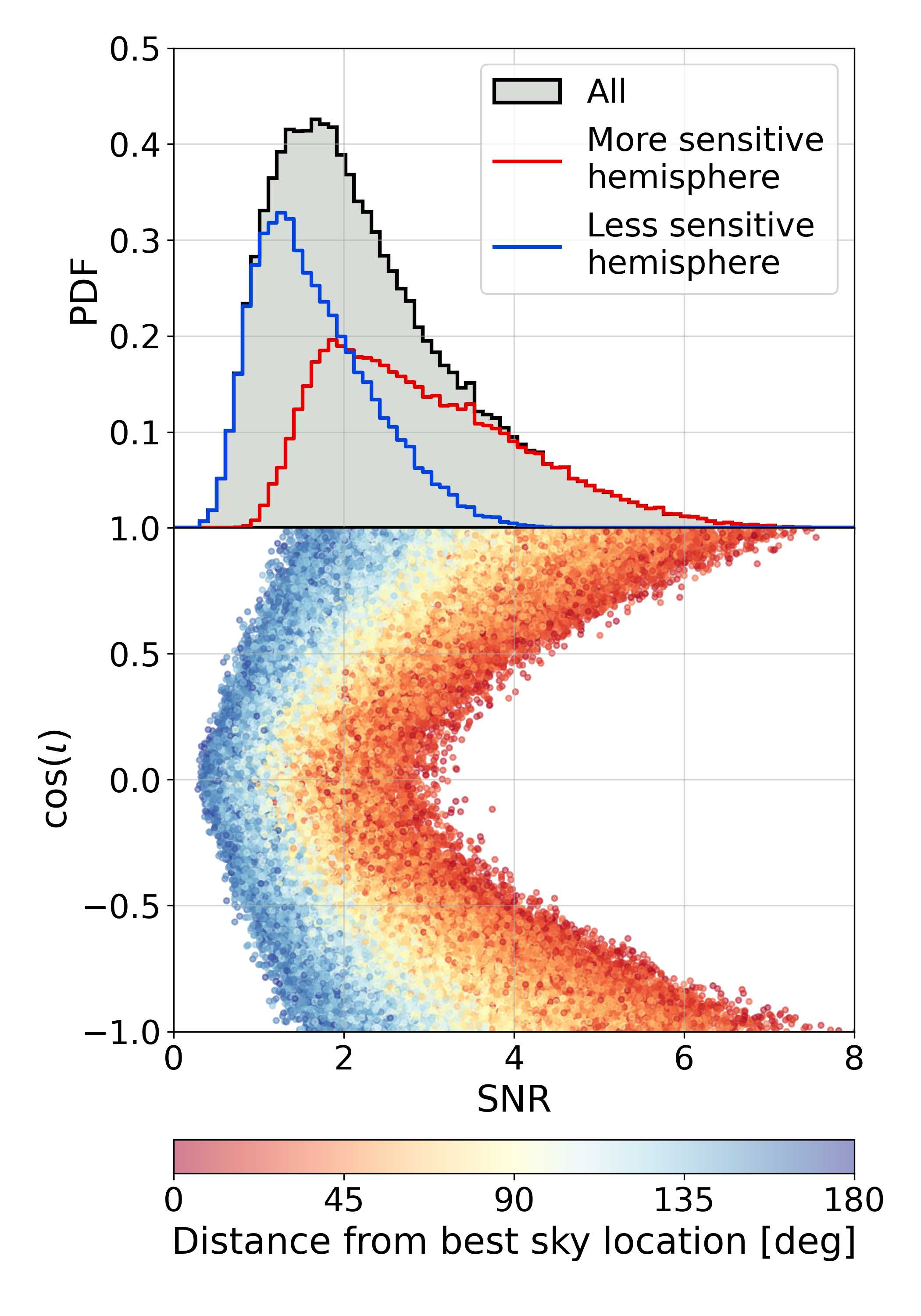}
 \caption{Distribution of SNRs for a particular source. The top panel shows the overall distribution (gray), along with histograms restricted to the half of the sky centered on the most sensitive sky location (red), and the antipodal point (blue). The bottom panel shows how SNR correlates with the orbital inclination ($\iota$). The coloring indicates each source's angular distance from the most sensitive sky location (see black cross in Fig.~\ref{fig:snr_sky}).}
 \label{fig:snr_dist}
\end{figure}

The top panel of Figure \ref{fig:snr_dist} shows the distribution of SNRs for this particular source, marginalized over all external parameters. There is a significant variance in the SNR, with the 95\% credible interval ranging from 0.8 to 5.1. We also show the distribution restricted to the more and less sensitive hemisphere, which we define relative to the most sensitive sky pixel marked with a black cross in Fig.~\ref{fig:snr_sky}. This shows that the sky location significantly contributes to the variance of the overall distribution. While most other extrinsic parameters do not show a clear correlation with SNR, the well-known effect of the inclination angle ($\iota$) appears here as well. This is evident from the strong dependence of the GW amplitude on $\iota$ (see e.g.~Eq.~(13-14) of \citealt{nanograv_11_cw}). The bottom panel of Figure \ref{fig:snr_dist} shows the 2-dimensional distribution of $\cos \iota$ and SNR. Sources with face-on orientation ($\cos \iota = \pm 1$) produce significantly higher SNR values compared to edge-on systems ($\cos \iota = 0$). We also color-code the points on this panel with the source's angular distance from the most sensitive sky location marked in Fig.~\ref{fig:snr_sky}. We can see that orbital inclination and sky location represent the majority of the SNR variance. This means that PTAs are more likely to first see an individual SMBHB with approximately face-on orientation, located in the part of the sky where our PTA is most sensitive towards.

\subsection{\label{ssec:best_sources}Expected properties of the highest-SNR SMBHB}
\begin{figure}[htb]
 \centering
   \includegraphics[width=\columnwidth]{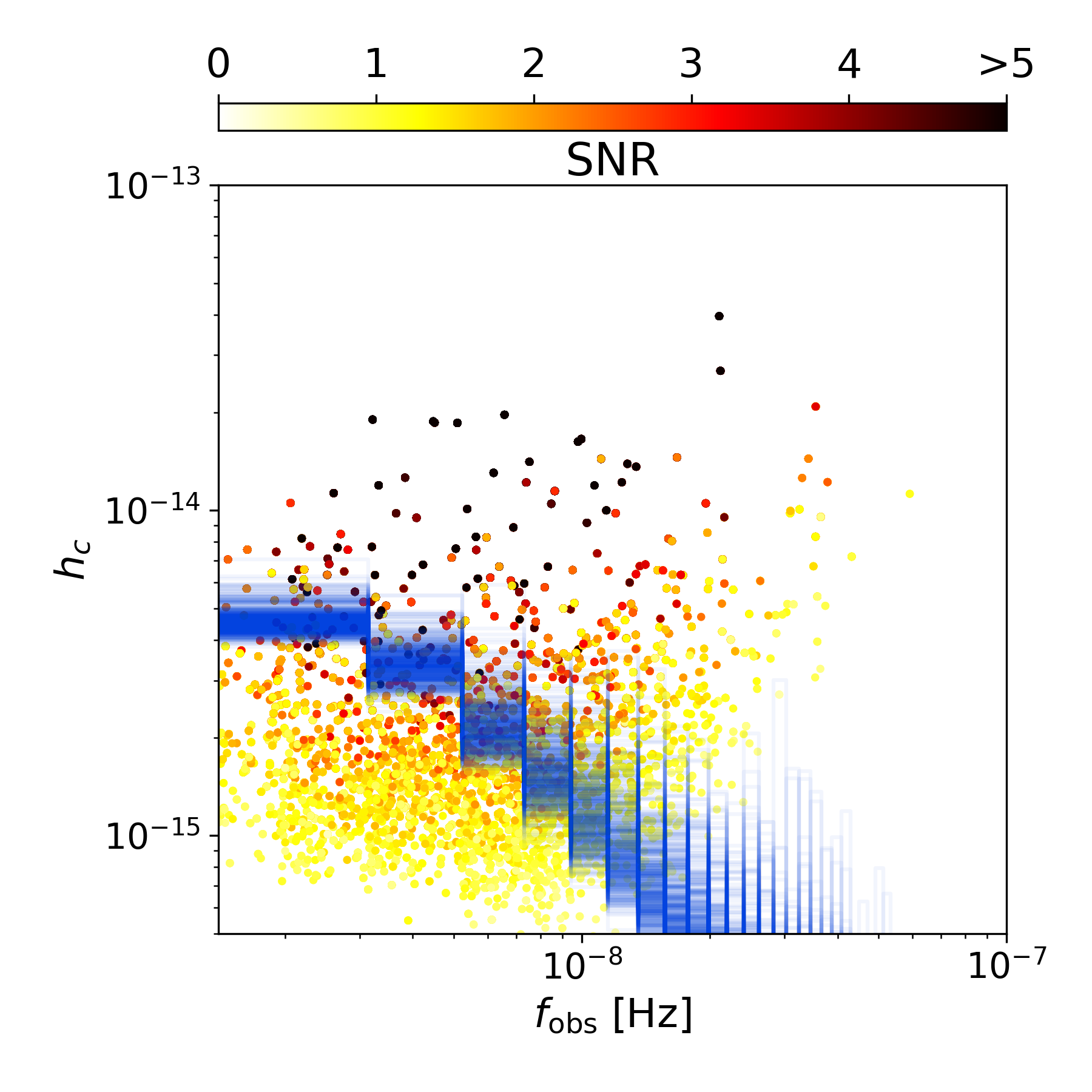}
 \caption{Location of the loudest sources on the $h_c$--$f_{\rm obs}$ plane over 500 realizations of the binary population with 100 random extrinsic parameters each (50 thousand in total). The color of each dot indicates the SNR of the loudest source in the particular realization. Note that the color scale was capped at SNR=5 for better visibility. We also show the GWB spectra for each realization in blue. Sources can have significant SNRs even if they are below the characteristic strain spectrum for the GWB if they have favorable extrinsic parameters (e.g.~sky locations and inclinations).}
 \label{fig:best_sources_spectrum}
\end{figure}

We are interested in the expected properties of the CW signal that will first be detectable. To find those, we create 50 thousand realizations of a realistic GWB made up from 500 realizations of the SMBHB population with 100 different random extrinsic parameters each. We use a simulated PTA based on the NANOGrav 12.5-year dataset extended to a 15 year timespan \citep{Astro4Cast}. Thus the following results can be treated as pessimistic predictions for the upcoming NANOGrav 15-year dataset \citep{nanograv_15_data}, since we do not take into account the improved timing precision and the addition of new pulsars. For each realization we find the individual binary with the highest SNR looking through all the frequency bins. Figure \ref{fig:best_sources_spectrum} shows the location of these loudest sources on the $h_c$--$f_{\rm obs}$ plane, color-coded with their corresponding SNR. We also show the spectrum of the GWB for all realizations. We can see that the loudest sources are concentrated at moderate frequencies. The lack of high-SNR sources at higher frequencies is due to the white noise in the dataset, which has $h_c \sim f_{\rm obs}^{3/2}$. Note that a large fraction of the loudest sources lie below the characteristic strain spectrum of the GWB, which act as a noise source here. This is possible due to the fact that the SNR not only depends on $h_c$, but also on extrinsic parameters as we have seen in Section \ref{ssec:snr_variance}. Selecting the loudest source over all frequencies preferentially selects sources with the most favorable sky locations, inclinations, etc, thus resulting in significant SNRs even if the source has $h_c<h_c^{\rm GWB}$.

\begin{figure*}[htb]
 \centering
   \includegraphics[width=0.9\textwidth]{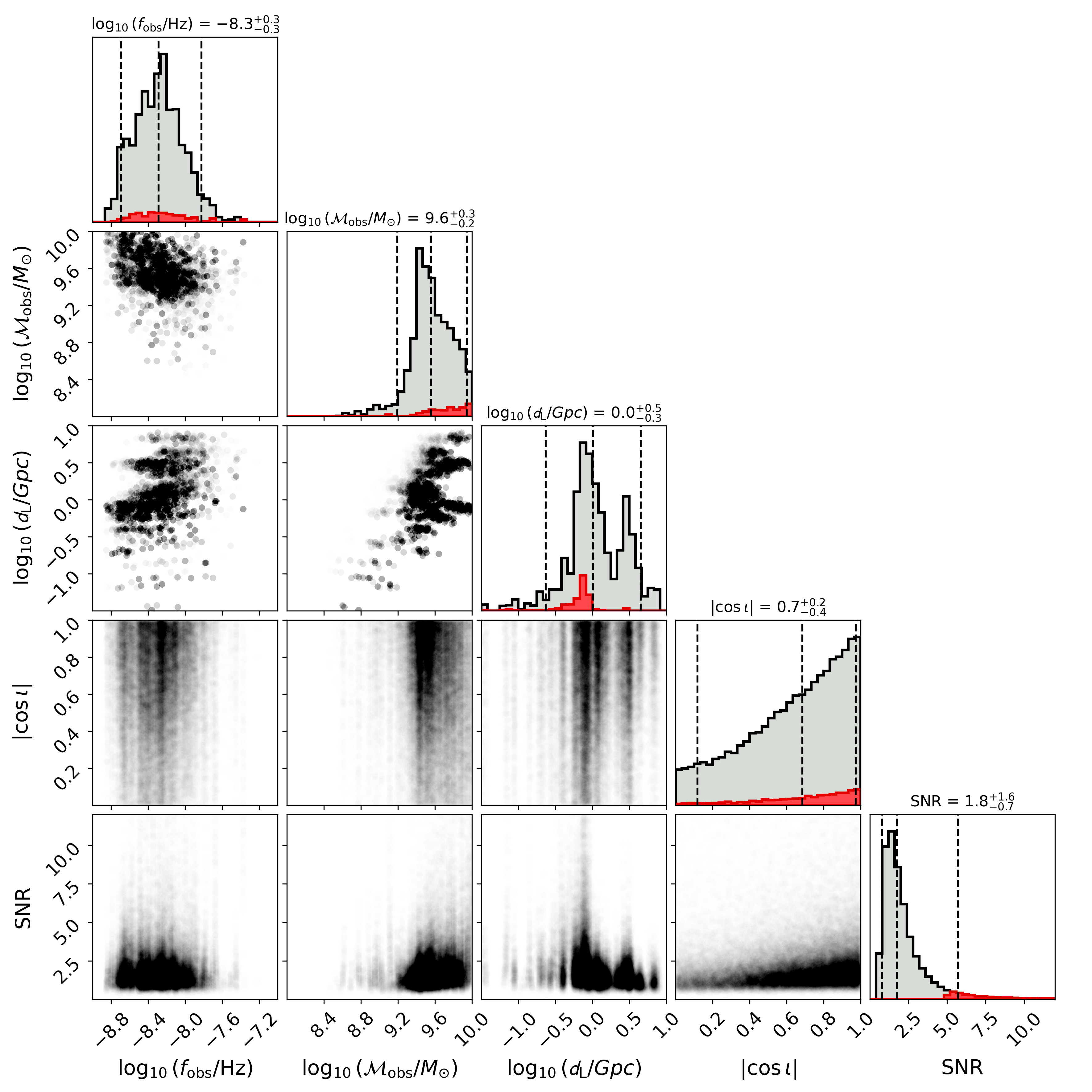}
 \caption{Distribution of the detector-frame GW frequency and chirp mass, luminosity distance, inclination angle, and the SNR of the loudest source over 500 binary population realizations with 100 realization each (50 thousand in total). Dashed lines and quoted values correspond to the 5th/50th/95th percentiles. Gray histograms show all realizations, while red histograms show only those where the SNR is larger than 5 ($\sim8$\% of all realizations).}
 \label{fig:best_params_corner}
\end{figure*}

Figure \ref{fig:best_params_corner} shows the 1 and 2-dimensional marginal distributions of the detector-frame GW frequency and chirp mass, luminosity distance, inclination angle, and the SNR for the loudest sources over 50 thousand realizations. The median $f_{\rm obs}$ is $\sim6$ nHz, the median chirp mass is $\sim5\times10^9 \ M_{\odot}$, and the median luminosity distance is $\sim1.5$ Gpc. This suggests that we will most likely first see a CW source at moderate frequencies with a very high chirp mass at a considerable distance. This is in agreement with the findings of \citet{rosado_expected_properties} and \citet{Luke_paper2}. The distribution of $\cos \iota$ is significantly different from the flat distribution corresponding to isotropic inclination distribution. This is a well-known selection effect due to the increased SNR for face-on systems (see Fig.~\ref{fig:snr_dist}). The probability of the loudest source being within 30$^\circ$ of face-on is $\sim26$\% for the whole population and $\sim38$\% for the SNR $>5$ subpopulation (compare with $\sim13$\% for the isotropic distribution).

We can also see on Figure \ref{fig:best_params_corner} that the median SNR is 1.8, which is not expected to be detectable. However, there is about a 6\% chance to get a source with SNR $>5$. Depending on the details of the detection algorithm used, those might be detectable, which is an interesting prospect for the search for individual binaries in the upcoming NANOGrav 15-year dataset \citep{nanograv_15_data}. We also show histograms for these high-SNR sources in red. We can see that they largely follow the same distribution for $f_{\rm obs}$, $\mathcal{M}$, and $d_{\rm L}$ as all realizations. The same analysis extended to a 20 year timespan yields similar $f_{\rm obs}$, $\mathcal{M}$, and $d_{\rm L}$ values, a median SNR of 4.5, and SNR $>5$ in about 41\% of the realizations.

Note that another reason why these predictions are pessimistic is that we only check the binary with the highest $h_c$ in each bin. In principle, binaries with lower $h_c$ can end up producing the highest SNR signals. This means that by not taking those into account, our SNR distributions are biased low. On the other hand, these results do not marginalize over the uncertainty in the astrophysical models used for the SMBHB populations, which could have a large effect on these results. We will incorporate these additional details in a future study to provide more robust predictions for future PTA detection prospects.


\section{\label{sec:conclusion}Conclusion and future work}

In this paper we presented a new approach to efficiently simulating realistic PTA datasets by modeling the signal as a combination of an isotropic Gaussian GWB and a few of the brightest binaries modeled individually. This produces datasets consistent with the naive method of directly modeling all binaries in a fraction of the time. We used this simulation technique to explore various properties of realistic PTA datasets. We show that the datasets are dominated by a small number of binaries at high frequencies, resulting in the well-known lack of GW power compared to a simple power-law model (see e.g.~\citealt{Sesana_high_f_discrepancy}). We test these datasets for statistical isotropy and find that they can be made isotropic by removing the few brightest binaries at all except the highest frequencies. We calculate the mean and the variance of the spatial correlations in our simulated datasets and find good agreement with analytical results presented in \citet{BruceAllen_HD_variance}. 

Our methodology also allows for calculating SNRs of any binary in our datasets. We use that to calculate the distribution of the SNR of the brightest source over realizations for a simulated PTA based on the NANOGrav 12.5-year pulsars with time spans extended to 15 years. These calculations account for both pulsar noise and the confusion noise from the GWB. We find that the brightest binary tends to have moderate GW frequency (few times the inverse of the observational timespan), high chirp mass, large distance, and nearly face-on orientation. The median SNR of the brightest source is 1.8, and about 6\% of the realizations produce a source with SNR higher than 5. These SNR values are pessimistic estimates for the upcoming NANOGrav 15-year dataset \citep{nanograv_15_data}, since they do not take into account any potential improvements of the timing model solutions or the addition of new pulsars relative to the 12.5-year dataset. However, they rely on a particular model of the SMBHB population, which might introduce biases. If we further increase the observing timespan to 20 years, the fraction of realizations producing a binary with SNR greater than 5 increases to about 41\%, and the median SNR increases to 4.5. We also calculate the SNR distribution for a given binary over different values of their extrinsic parameters (sky location, inclination and polarization angle, initial phase). We find that these parameters can have a significant effect on the SNR. The sky location and the inclination angle are particularly impactful parameters, as they can change the SNR by a factor of $2-4$.

The methods presented in this paper will continue to be useful tools to understand the interplay between the stochastic background and individual sources. In the future, we plan to incorporate different models of the SMBHB population, allowing us to produce simulated datasets using different assumptions about SMBHB formation and evolution. The software used for this paper was made to be compatible with the \texttt{holodeck}\footnote{Publicly available at: \url{https://github.com/nanograv/holodeck}} software package \citep{holodeck_methods}, allowing us to use any SMBHB population model developed there. That will also allow us to make our detection prospect predictions marginalized over astrophysical uncertainties. In addition, we also plan to run PTA detection pipelines on realistic datasets produced by the methods presented here. In particular, analyzing such datasets with the \texttt{QuickCW}\footnote{Publicly available at: \url{https://github.com/bencebecsy/QuickCW}} pipeline \citep{QuickCW}, would allow us to make more accurate predictions than the simple SNR calculations presented in this paper. We also plan to analyze such datasets with \texttt{BayesHopper} to investigate how PTAs will be able to detecting multiple individual sources in the future \citep{BayesHopper}.

\begin{acknowledgments}
The authors thank Stephen Taylor and Nihan Pol for providing access to software used in \citet{Astro4Cast} to create simulated datasets based on measured noise properties of NANOGrav pulsars. We thank Chiara Mingarelli and Alberto Sesana for feedback on the manuscript. We also thank Bruce Allen and Joseph Romano for useful discussions about the variance of HD correlations. We appreciate the support of the NSF Physics Frontiers Center Award PFC-1430284 and the NSF Physics Frontiers Center Award PFC-2020265.
\end{acknowledgments}


\software{\texttt{healpy} \citep{healpy},  
          \texttt{HEALPix} \citep{healpix},
          \texttt{ENTERPRISE} \citep{enterprise},
          \texttt{corner} \citep{corner},
          \texttt{Numba} \citep{numba_paper, numba_zenodo}
          }

\appendix

\section{Variance of spatial correlations with different normalizations}
\label{sec:hd_variance_appendix}

In this appendix we explore how the variance of the HD correlations from a single binary depend on the way we normalize the correlation values. We reproduce some results of \citet{BruceAllen_HD_variance} and derive formulae for the case with a different normalization used in Section \ref{ssec:hd-variance}. We compare these results with simulations for a CW signal in a simulated PTA with an isotropic distribution of pulsars. Note that we follow the convention that the HD curve at zero separation is 1/2, while \citet{BruceAllen_HD_variance} normalizes to 1/3. Thus our results need to be multiplied by 2/3 when comparing with results in \citet{BruceAllen_HD_variance}.

\subsection{Without pulsar terms}

From Eq.~(A31) of \citet{BruceAllen_HD_variance} the mean HD correlation for a binary with a given inclination ($\iota$) and GW amplitude ($\mathcal{A}$) is:
\begin{equation}
    \mu = \frac{\mathcal{A}^2}{16} \left[ 1 + 6 \cos ^2 \iota + \cos ^4 \iota \right] \mu_u (\gamma),
\label{eq:mu}
\end{equation}
where $\mu_u (\gamma)$ is the mean HD curve (see Eq.~(D29) in \citealt{BruceAllen_HD_variance}). Neglecting pulsar terms the variance is:
\begin{multline}
    \sigma ^2 = \frac{\mathcal{A}^4}{256} \left[ 1 + 6 \cos ^2 \iota + \cos ^4 \iota \right]^2 \sigma_u^2 (\gamma) + \\
     \frac{\mathcal{A}^4}{512} \left[ \sin ^8 \iota \right] \sigma_c^2 (\gamma),
\end{multline}
where $\sigma_u$ and $\sigma_c$ are defined in Eqs. (D37) and (F2) in \citet{BruceAllen_HD_variance} respectively.

We can average these over $\cos \iota \in [-1,1]$ to get (see Eq.~(A32) of \citealt{BruceAllen_HD_variance}):
\begin{equation}
    \mu_{\rm average} = \frac{\mathcal{A}^2}{5} \mu_u (\gamma),
\end{equation}
and:
\begin{equation}
    \sigma^2_{\rm average} = \frac{71 \mathcal{A}^4}{1260} \sigma_u^2 (\gamma) + \frac{\mathcal{A}^4}{1260} \sigma_c^2 (\gamma).
\end{equation}

We normalize these so that $\mu_{\rm average} = \mu_u$, which implies $\mathcal{A}^2=5$ and:
\begin{equation}
    \sigma_{\rm average} = \sqrt{\frac{355}{252} \sigma_u^2 (\gamma) + \frac{5}{252} \sigma_c^2 (\gamma)}.
\label{eq:global_norm}
\end{equation}

This is appropriate if we normalize the correlation values globally over all realizations. However, if we normalize the correlations realization by realization, then we need to normalize Eq.~(\ref{eq:mu}) before averaging. We thus require $\mathcal{A}^2/16 \left[ 1 + 6 \cos ^2 \iota + \cos ^4 \iota \right] = 1$, which implies:
\begin{equation}
    \sigma ^2 = \sigma_u^2 (\gamma) + \frac{\sin ^8 \iota}{2 \left[ 1 + 6 \cos ^2 \iota + \cos ^4 \iota \right]^2} \sigma_c^2 (\gamma). 
\end{equation}

If we average this over $\cos \iota \in [-1,1]$, we get:
\begin{equation}
    \sigma_{\rm average} = \sqrt{ \sigma_u^2 (\gamma) + \frac{1}{4} \alpha \sigma_c^2 (\gamma)},
\label{eq:per_real_norm}
\end{equation}
where $\alpha=4+\pi-3\pi/\sqrt{2}$.

\begin{figure}[h!]
 \centering
   \includegraphics[width=\columnwidth]{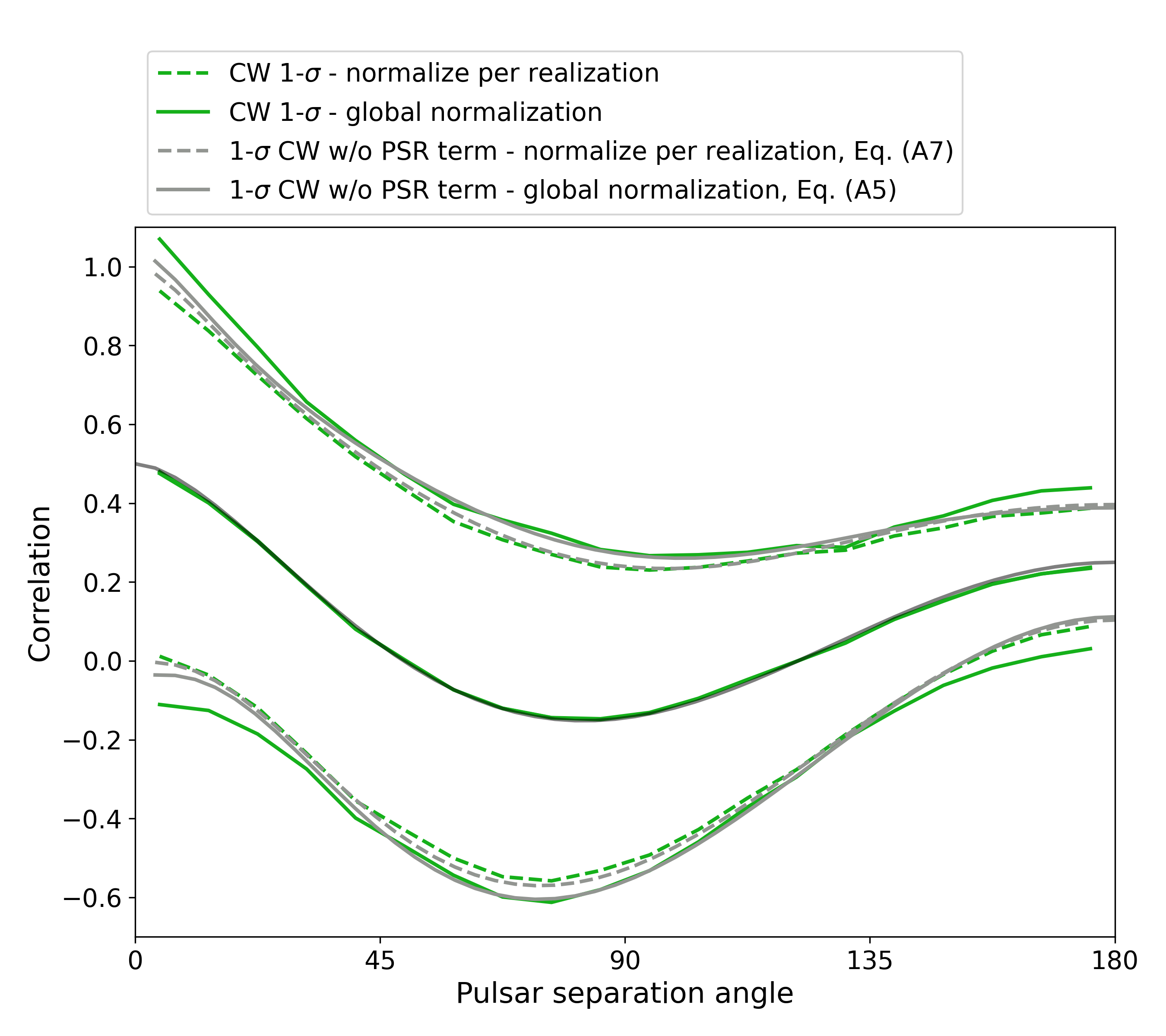}
 \caption{Variance of HD correlations from a CW simulation without pulsar terms and a simulated PTA with isotropically distributed pulsars. We also show the theoretical variance with two different averaging.}
 \label{fig:no_psr_term}
\end{figure}

Figure \ref{fig:no_psr_term} shows the comparison of simulation results with Eqs.~(\ref{eq:global_norm}) and (\ref{eq:per_real_norm}). The simulation normalized per realization seems to match Eq.~(\ref{eq:per_real_norm}) really well. The one normalized globally generally follows Eq.~(\ref{eq:global_norm}), but deviates at small and large angles.

\subsection{With pulsar terms}

If we include pulsar terms, the mean correlation is unchanged, but the variance becomes (cf.~Eq.~(B10) of \citealt{BruceAllen_HD_variance}):

\begin{multline}
    \sigma ^2 = \frac{\mathcal{A}^4}{256} \left[ 1 + 6 \cos ^2 \iota + \cos ^4 \iota \right]^2 \sigma_u^2 (\gamma) - \\
    \frac{3 \mathcal{A}^4}{512} \left[ \sin ^8 \iota \right] \sigma_p^2 (\gamma) + \\
    \mathcal{A}^4 \left( \frac{3}{512} \left[ 1 + 6 \cos ^2 \iota + \cos ^4 \iota \right]^2 + \frac{5}{1024} \left[ \sin ^8 \iota \right] \right) \sigma_c^2 (\gamma).
    \label{eq:psr_term}
\end{multline}
where $\sigma_p$ is defined in Eq. (E8) in \citet{BruceAllen_HD_variance}.

Averaging this over $\cos \iota \in [-1,1]$ gives (see Eq.~(B11) of \citealt{BruceAllen_HD_variance}):
\begin{equation}
    \sigma ^2 = \frac{71 \mathcal{A}^4}{1260} \sigma_u^2 (\gamma) - \frac{3 \mathcal{A}^4}{1260} \sigma_p^2 (\gamma) + \frac{109 \mathcal{A}^4}{1260} \sigma_c^2 (\gamma).
\end{equation}

Plugging in $\mathcal{A}^4=25$ gives:
\begin{equation}
    \sigma_{\rm average} = \sqrt{ \frac{355}{252} \sigma_u^2 (\gamma) - \frac{5}{84} \sigma_p^2 (\gamma) + \frac{545}{252} \sigma_c^2 (\gamma)}.
    \label{eq:globan_with_psr_term}
\end{equation}

This is the average standard deviation if we normalize correlations globally, over many realizations. Alternatively, we can express $\mathcal{A}$ from Eq.~(\ref{eq:mu}) and plug into Eq.~(\ref{eq:psr_term}) before averaging. This gives:
\begin{multline}
    \sigma ^2 = \sigma_u^2 (\gamma) - \frac{3 \sin ^8 \iota}{2 \left[ 1 + 6 \cos ^2 \iota + \cos ^4 \iota \right]^2} \sigma_p^2 (\gamma) + \\
    \left( \frac{3}{2} + \frac{5 \sin ^8 \iota}{4 \left[ 1 + 6 \cos ^2 \iota + \cos ^4 \iota \right]^2} \right) \sigma_c^2 (\gamma).
\end{multline}

Averaging over $\cos \iota \in [-1,1]$ gives:
\begin{equation}
    \sigma_{\rm average} = \sqrt{\sigma_u^2 (\gamma) - \frac{3}{4} \alpha \sigma_p^2 (\gamma) + \left( \frac{3}{2} + \frac{5}{8} \alpha \right) \sigma_c^2 (\gamma)}.
    \label{eq:per_real_with_psr_term}
\end{equation}

\begin{figure}[htb]
 \centering
   \includegraphics[width=\columnwidth]{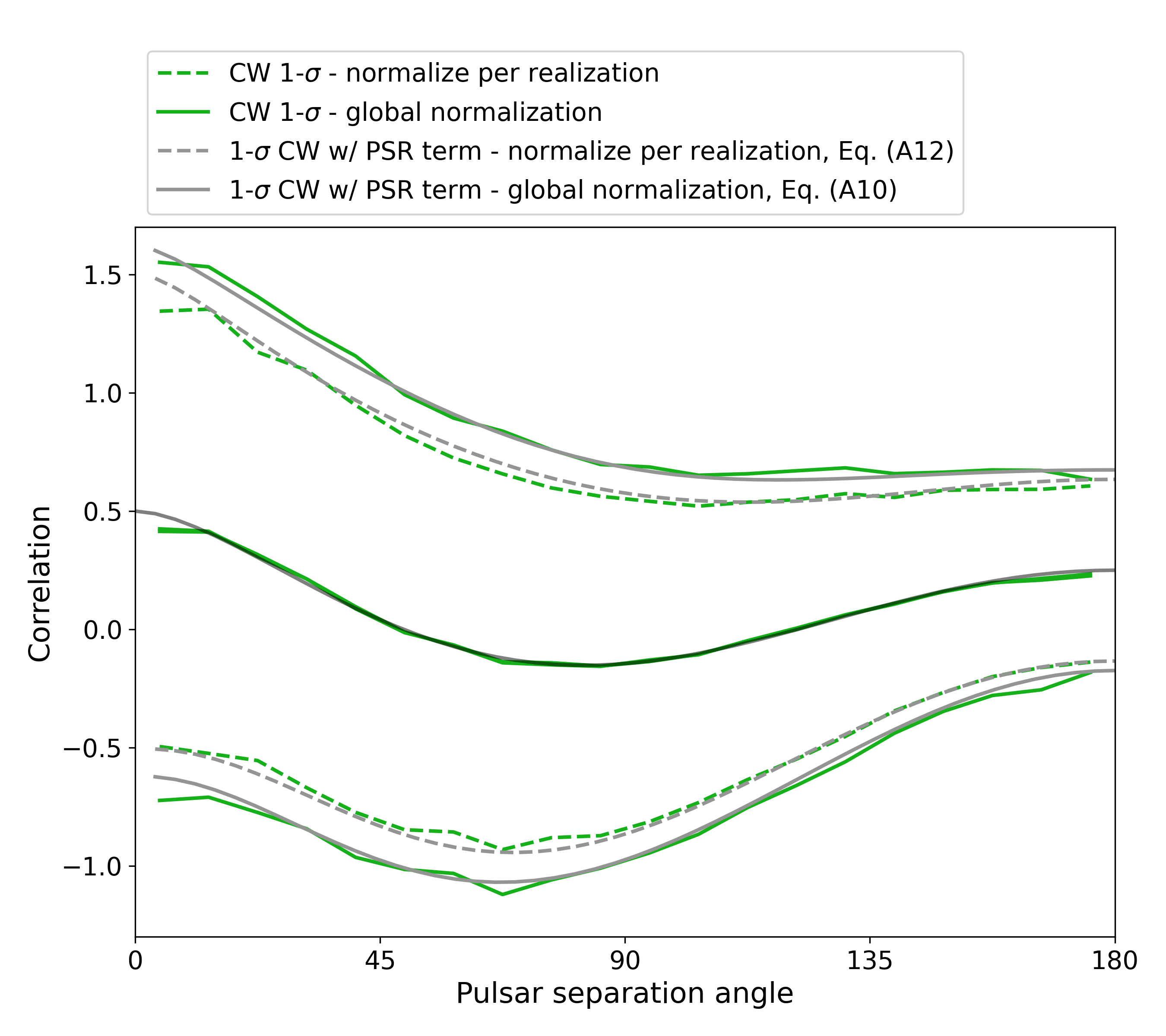}
 \caption{Variance of HD correlations from a CW simulation with pulsar terms and a simulated PTA with isotropically distributed pulsars. We also show the theoretical variance with two different averaging.}
 \label{fig:psr_term}
\end{figure}

Figure \ref{fig:psr_term} shows simulation results (both with global normalization and per realization normalization) and Eqs. (\ref{eq:globan_with_psr_term}) and (\ref{eq:per_real_with_psr_term}). We can see that both match well with the appropriate simulation.

\bibliography{realisticGWB}{}

\providecommand{\noopsort}[1]{}\providecommand{\singleletter}[1]{#1}%
\begin{thebibliography}{}
\expandafter\ifx\csname natexlab\endcsname\relax\def\natexlab#1{#1}\fi
\providecommand{\url}[1]{\href{#1}{#1}}
\providecommand{\dodoi}[1]{doi:~\href{http://doi.org/#1}{\nolinkurl{#1}}}
\providecommand{\doeprint}[1]{\href{http://ascl.net/#1}{\nolinkurl{http://ascl.net/#1}}}
\providecommand{\doarXiv}[1]{\href{https://arxiv.org/abs/#1}{\nolinkurl{https://arxiv.org/abs/#1}}}

\bibitem[{{Aggarwal} {et~al.}(2019){Aggarwal}, {Arzoumanian}, {Baker},
  {Brazier}, {Brinson}, {Brook}, {Burke-Spolaor}, {Chatterjee}, {Cordes},
  {Cornish}, {Crawford}, {Crowter}, {Cromartie}, {DeCesar}, {Demorest},
  {Dolch}, {Ellis}, {Ferdman}, {Ferrara}, {Fonseca}, {Garver-Daniels},
  {Gentile}, {Hazboun}, {Holgado}, {Huerta}, {Islo}, {Jennings}, {Jones},
  {Jones}, {Kaiser}, {Kaplan}, {Kelley}, {Key}, {Lam}, {Lazio}, {Levin},
  {Lorimer}, {Luo}, {Lynch}, {Madison}, {McLaughlin}, {McWilliams},
  {Mingarelli}, {Ng}, {Nice}, {Pennucci}, {Pol}, {Ransom}, {Ray}, {Siemens},
  {Simon}, {Spiewak}, {Stairs}, {Stinebring}, {Stovall}, {Swiggum}, {Taylor},
  {Turner}, {Vallisneri}, {van Haasteren}, {Vigeland}, {Witt}, {Zhu}, \&
  {NANOGrav Collaboration}}]{nanograv_11_cw}
{Aggarwal}, K., {Arzoumanian}, Z., {Baker}, P.~T., {et~al.} 2019, \apj, 880,
  116, \dodoi{10.3847/1538-4357/ab2236}

\bibitem[{{Alam} {et~al.}(2021){Alam}, {Arzoumanian}, {Baker}, {Blumer},
  {Bohler}, {Brazier}, {Brook}, {Burke-Spolaor}, {Caballero}, {Camuccio},
  {Chamberlain}, {Chatterjee}, {Cordes}, {Cornish}, {Crawford}, {Cromartie},
  {Decesar}, {Demorest}, {Dolch}, {Ellis}, {Ferdman}, {Ferrara}, {Fiore},
  {Fonseca}, {Garcia}, {Garver-Daniels}, {Gentile}, {Good}, {Gusdorff},
  {Halmrast}, {Hazboun}, {Islo}, {Jennings}, {Jessup}, {Jones}, {Kaiser},
  {Kaplan}, {Kelley}, {Key}, {Lam}, {Lazio}, {Lorimer}, {Luo}, {Lynch},
  {Madison}, {Maraccini}, {McLaughlin}, {Mingarelli}, {Ng}, {Nguyen}, {Nice},
  {Pennucci}, {Pol}, {Ramette}, {Ransom}, {Ray}, {Shapiro-Albert}, {Siemens},
  {Simon}, {Spiewak}, {Stairs}, {Stinebring}, {Stovall}, {Swiggum}, {Taylor},
  {Tripepi}, {Vallisneri}, {Vigeland}, {Witt}, {Zhu}, \& {Nanograv
  Collaboration}}]{nanograv_12p5_data}
{Alam}, M.~F., {Arzoumanian}, Z., {Baker}, P.~T., {et~al.} 2021, \apjs, 252, 4,
  \dodoi{10.3847/1538-4365/abc6a0}

\bibitem[{{Alam} {et~al.}(in prep.){Alam}, {Arzoumanian}, {Baker}, {Blumer},
  {Bohler}, {Brazier}, {Brook}, {Burke-Spolaor}, {Caballero}, {Camuccio},
  {Chamberlain}, {Chatterjee}, {Cordes}, {Cornish}, {Crawford}, {Cromartie},
  {Decesar}, {Demorest}, {Dolch}, {Ellis}, {Ferdman}, {Ferrara}, {Fiore},
  {Fonseca}, {Garcia}, {Garver-Daniels}, {Gentile}, {Good}, {Gusdorff},
  {Halmrast}, {Hazboun}, {Islo}, {Jennings}, {Jessup}, {Jones}, {Kaiser},
  {Kaplan}, {Kelley}, {Key}, {Lam}, {Lazio}, {Lorimer}, {Luo}, {Lynch},
  {Madison}, {Maraccini}, {McLaughlin}, {Mingarelli}, {Ng}, {Nguyen}, {Nice},
  {Pennucci}, {Pol}, {Ramette}, {Ransom}, {Ray}, {Shapiro-Albert}, {Siemens},
  {Simon}, {Spiewak}, {Stairs}, {Stinebring}, {Stovall}, {Swiggum}, {Taylor},
  {Tripepi}, {Vallisneri}, {Vigeland}, {Witt}, {Zhu}, \& {Nanograv
  Collaboration}}]{nanograv_15_data}
---. in prep.

\bibitem[{Ali-Ha\"{\i}moud {et~al.}(2020)Ali-Ha\"{\i}moud, Smith, \&
  Mingarelli}]{Chiara_anisotropy_2020}
Ali-Ha\"{\i}moud, Y., Smith, T.~L., \& Mingarelli, C. M.~F. 2020, Phys. Rev. D,
  102, 122005, \dodoi{10.1103/PhysRevD.102.122005}

\bibitem[{{Allen}(2022)}]{BruceAllen_HD_variance}
{Allen}, B. 2022, arXiv e-prints, arXiv:2205.05637.
\newblock \doarXiv{2205.05637}

\bibitem[{{Allen} \& {Romano}(2022)}]{allen_romano_cosmic_variance}
{Allen}, B., \& {Romano}, J.~D. 2022, arXiv e-prints, arXiv:2208.07230.
\newblock \doarXiv{2208.07230}

\bibitem[{{Antoniadis} {et~al.}(2022){Antoniadis}, {Arzoumanian}, {Babak},
  {Bailes}, {Bak Nielsen}, {Baker}, {Bassa}, {B{\'e}csy}, {Berthereau},
  {Bonetti}, {Brazier}, {Brook}, {Burgay}, {Burke-Spolaor}, {Caballero},
  {Casey-Clyde}, {Chalumeau}, {Champion}, {Charisi}, {Chatterjee}, {Chen},
  {Cognard}, {Cordes}, {Cornish}, {Crawford}, {Cromartie}, {Crowter}, {Dai},
  {DeCesar}, {Demorest}, {Desvignes}, {Dolch}, {Drachler}, {Falxa}, {Ferrara},
  {Fiore}, {Fonseca}, {Gair}, {Garver-Daniels}, {Goncharov}, {Good}, {Graikou},
  {Guillemot}, {Guo}, {Hazboun}, {Hobbs}, {Hu}, {Islo}, {Janssen}, {Jennings},
  {Johnson}, {Jones}, {Kaiser}, {Kaplan}, {Karuppusamy}, {Keith}, {Kelley},
  {Kerr}, {Key}, {Kramer}, {Lam}, {Lamb}, {Lazio}, {Lee}, {Lentati}, {Liu},
  {Luo}, {Lynch}, {Lyne}, {Madison}, {Main}, {Manchester}, {McEwen}, {McKee},
  {McLaughlin}, {Mickaliger}, {Mingarelli}, {Ng}, {Nice}, {Os{\l}owski},
  {Parthasarathy}, {Pennucci}, {Perera}, {Perrodin}, {Petiteau}, {Pol},
  {Porayko}, {Possenti}, {Ransom}, {Ray}, {Reardon}, {Russell}, {Samajdar},
  {Sampson}, {Sanidas}, {Sarkissian}, {Schmitz}, {Schult}, {Sesana},
  {Shaifullah}, {Shannon}, {Shapiro-Albert}, {Siemens}, {Simon}, {Smith},
  {Speri}, {Spiewak}, {Stairs}, {Stappers}, {Stinebring}, {Swiggum}, {Taylor},
  {Theureau}, {Tiburzi}, {Vallisneri}, {van der Wateren}, {Vecchio},
  {Verbiest}, {Vigeland}, {Wahl}, {Wang}, {Wang}, {Wang}, {Witt}, {Zhang}, \&
  {Zhu}}]{IPTA_DR2_GWB}
{Antoniadis}, J., {Arzoumanian}, Z., {Babak}, S., {et~al.} 2022, \mnras, 510,
  4873, \dodoi{10.1093/mnras/stab3418}

\bibitem[{{Arzoumanian} {et~al.}(2020){Arzoumanian}, {Baker}, {Blumer},
  {B{\'e}csy}, {Brazier}, {Brook}, {Burke-Spolaor}, {Chatterjee}, {Chen},
  {Cordes}, {Cornish}, {Crawford}, {Cromartie}, {Decesar}, {Demorest}, {Dolch},
  {Ellis}, {Ferrara}, {Fiore}, {Fonseca}, {Garver-Daniels}, {Gentile}, {Good},
  {Hazboun}, {Holgado}, {Islo}, {Jennings}, {Jones}, {Kaiser}, {Kaplan},
  {Kelley}, {Key}, {Laal}, {Lam}, {Lazio}, {Lorimer}, {Luo}, {Lynch},
  {Madison}, {McLaughlin}, {Mingarelli}, {Ng}, {Nice}, {Pennucci}, {Pol},
  {Ransom}, {Ray}, {Shapiro-Albert}, {Siemens}, {Simon}, {Spiewak}, {Stairs},
  {Stinebring}, {Stovall}, {Sun}, {Swiggum}, {Taylor}, {Turner}, {Vallisneri},
  {Vigeland}, {Witt}, \& {Nanograv Collaboration}}]{NANOGrav_12p5_gwb}
{Arzoumanian}, Z., {Baker}, P.~T., {Blumer}, H., {et~al.} 2020, \apjl, 905,
  L34, \dodoi{10.3847/2041-8213/abd401}

\bibitem[{{Arzoumanian} {et~al.}(2021){Arzoumanian}, {Baker}, {Blumer},
  {B{\'e}csy}, {Brazier}, {Brook}, {Burke-Spolaor}, {Charisi}, {Chatterjee},
  {Chen}, {Cordes}, {Cornish}, {Crawford}, {Cromartie}, {Decesar}, {Demorest},
  {Dolch}, {Ellis}, {Ferrara}, {Fiore}, {Fonseca}, {Garver-Daniels}, {Gentile},
  {Good}, {Hazboun}, {Holgado}, {Islo}, {Jennings}, {Jones}, {Kaiser},
  {Kaplan}, {Kelley}, {Key}, {Laal}, {Lam}, {Lazio}, {Lee}, {Lorimer}, {Luo},
  {Lynch}, {Madison}, {McLaughlin}, {Mingarelli}, {Mitridate}, {Ng}, {Nice},
  {Pennucci}, {Pol}, {Ransom}, {Ray}, {Shapiro-Albert}, {Siemens}, {Simon},
  {Spiewak}, {Stairs}, {Stinebring}, {Stovall}, {Sun}, {Swiggum}, {Taylor},
  {Turner}, {Vallisneri}, {Vigeland}, {Witt}, {Zurek}, \& {Nanograv
  Collaboration}}]{NANOGrav_12p5_phase_transition}
---. 2021, \prl, 127, 251302, \dodoi{10.1103/PhysRevLett.127.251302}

\bibitem[{{B{\'e}csy} \& {Cornish}(2020)}]{BayesHopper}
{B{\'e}csy}, B., \& {Cornish}, N.~J. 2020, Classical and Quantum Gravity, 37,
  135011, \dodoi{10.1088/1361-6382/ab8bbd}

\bibitem[{{B{\'e}csy} {et~al.}(2022){B{\'e}csy}, {Cornish}, \&
  {Digman}}]{QuickCW}
{B{\'e}csy}, B., {Cornish}, N.~J., \& {Digman}, M.~C. 2022, arXiv e-prints,
  arXiv:2204.07160.
\newblock \doarXiv{2204.07160}

\bibitem[{{Blasi} {et~al.}(2021){Blasi}, {Brdar}, \&
  {Schmitz}}]{KaiSchmitz_cosmic_stings}
{Blasi}, S., {Brdar}, V., \& {Schmitz}, K. 2021, \prl, 126, 041305,
  \dodoi{10.1103/PhysRevLett.126.041305}

\bibitem[{{Burke-Spolaor} {et~al.}(2019){Burke-Spolaor}, {Taylor}, {Charisi},
  {Dolch}, {Hazboun}, {Holgado}, {Kelley}, {Lazio}, {Madison}, {McMann},
  {Mingarelli}, {Rasskazov}, {Siemens}, {Simon}, \& {Smith}}]{PTA_review}
{Burke-Spolaor}, S., {Taylor}, S.~R., {Charisi}, M., {et~al.} 2019, \aapr, 27,
  5, \dodoi{10.1007/s00159-019-0115-7}

\bibitem[{{Chen} {et~al.}(2021){Chen}, {Caballero}, {Guo}, {Chalumeau}, {Liu},
  {Shaifullah}, {Lee}, {Babak}, {Desvignes}, {Parthasarathy}, {Hu}, {van der
  Wateren}, {Antoniadis}, {Bak Nielsen}, {Bassa}, {Berthereau}, {Burgay},
  {Champion}, {Cognard}, {Falxa}, {Ferdman}, {Freire}, {Gair}, {Graikou},
  {Guillemot}, {Jang}, {Janssen}, {Karuppusamy}, {Keith}, {Kramer}, {Liu},
  {Lyne}, {Main}, {McKee}, {Mickaliger}, {Perera}, {Perrodin}, {Petiteau},
  {Porayko}, {Possenti}, {Samajdar}, {Sanidas}, {Sesana}, {Speri}, {Stappers},
  {Theureau}, {Tiburzi}, {Vecchio}, {Verbiest}, {Wang}, {Wang}, \&
  {Xu}}]{EPTA_dr2_gwb}
{Chen}, S., {Caballero}, R.~N., {Guo}, Y.~J., {et~al.} 2021, \mnras, 508, 4970,
  \dodoi{10.1093/mnras/stab2833}

\bibitem[{Cornish \& Romano(2015)}]{PhysRevD.92.042001}
Cornish, N.~J., \& Romano, J.~D. 2015, Phys. Rev. D, 92, 042001,
  \dodoi{10.1103/PhysRevD.92.042001}

\bibitem[{{Cornish} \& {Sampson}(2016)}]{Neil_Laura_finite_number}
{Cornish}, N.~J., \& {Sampson}, L. 2016, \prd, 93, 104047,
  \dodoi{10.1103/PhysRevD.93.104047}

\bibitem[{{Cornish} \& {Sesana}(2013)}]{Neil_CW_HD}
{Cornish}, N.~J., \& {Sesana}, A. 2013, Classical and Quantum Gravity, 30,
  224005, \dodoi{10.1088/0264-9381/30/22/224005}

\bibitem[{{Cornish} \& {van Haasteren}(2014)}]{Neil_anisotropic_search}
{Cornish}, N.~J., \& {van Haasteren}, R. 2014, arXiv e-prints, arXiv:1406.4511.
\newblock \doarXiv{1406.4511}

\bibitem[{{De Luca} {et~al.}(2021){De Luca}, {Franciolini}, \&
  {Riotto}}]{primordial_bh_gwb}
{De Luca}, V., {Franciolini}, G., \& {Riotto}, A. 2021, \prl, 126, 041303,
  \dodoi{10.1103/PhysRevLett.126.041303}

\bibitem[{Di~Matteo {et~al.}(2019)Di~Matteo, King, \& Cornish}]{Neil_book}
Di~Matteo, T., King, A., \& Cornish, N.~J. 2019, Black hole formation and
  growth (Springer)

\bibitem[{Ellis {et~al.}(2020)Ellis, Vallisneri, Taylor, \& Baker}]{enterprise}
Ellis, J.~A., Vallisneri, M., Taylor, S.~R., \& Baker, P.~T. 2020, ENTERPRISE:
  Enhanced Numerical Toolbox Enabling a Robust PulsaR Inference SuitE, Zenodo,
  \dodoi{10.5281/zenodo.4059815}

\bibitem[{{Enoki} {et~al.}(2004){Enoki}, {Inoue}, {Nagashima}, \&
  {Sugiyama}}]{smbh_gwb_enoki}
{Enoki}, M., {Inoue}, K.~T., {Nagashima}, M., \& {Sugiyama}, N. 2004, \apj,
  615, 19, \dodoi{10.1086/424475}

\bibitem[{Foreman-Mackey(2016)}]{corner}
Foreman-Mackey, D. 2016, The Journal of Open Source Software, 1, 24,
  \dodoi{10.21105/joss.00024}

\bibitem[{{Goncharov} {et~al.}(2021){Goncharov}, {Shannon}, {Reardon}, {Hobbs},
  {Zic}, {Bailes}, {Cury{\l}o}, {Dai}, {Kerr}, {Lower}, {Manchester}, {Mandow},
  {Middleton}, {Miles}, {Parthasarathy}, {Thrane}, {Thyagarajan}, {Xue}, {Zhu},
  {Cameron}, {Feng}, {Luo}, {Russell}, {Sarkissian}, {Spiewak}, {Wang}, {Wang},
  {Zhang}, \& {Zhang}}]{PPTA_dr2_gwb}
{Goncharov}, B., {Shannon}, R.~M., {Reardon}, D.~J., {et~al.} 2021, \apjl, 917,
  L19, \dodoi{10.3847/2041-8213/ac17f4}

\bibitem[{{G{\'o}rski} {et~al.}(2005){G{\'o}rski}, {Hivon}, {Banday},
  {Wandelt}, {Hansen}, {Reinecke}, \& {Bartelmann}}]{healpix}
{G{\'o}rski}, K.~M., {Hivon}, E., {Banday}, A.~J., {et~al.} 2005, \apj, 622,
  759, \dodoi{10.1086/427976}

\bibitem[{Hajian \& Souradeep(2003)}]{kappa_test_2003}
Hajian, A., \& Souradeep, T. 2003, The Astrophysical Journal, 597, L5,
  \dodoi{10.1086/379757}

\bibitem[{{Hajian} \& {Souradeep}(2004)}]{kappa_test_formalism}
{Hajian}, A., \& {Souradeep}, T. 2004, arXiv e-prints, astro.
\newblock \doarXiv{astro-ph/0501001}

\bibitem[{{Hajian} {et~al.}(2005){Hajian}, {Souradeep}, \&
  {Cornish}}]{kappa_test_Neil}
{Hajian}, A., {Souradeep}, T., \& {Cornish}, N. 2005, \apjl, 618, L63,
  \dodoi{10.1086/427652}

\bibitem[{{Hellings} \& {Downs}(1983)}]{HD}
{Hellings}, R.~W., \& {Downs}, G.~S. 1983, \apjl, 265, L39,
  \dodoi{10.1086/183954}

\bibitem[{{Hotinli} {et~al.}(2019){Hotinli}, {Kamionkowski}, \&
  {Jaffe}}]{BipoSH_anisotropy}
{Hotinli}, S.~C., {Kamionkowski}, M., \& {Jaffe}, A.~H. 2019, The Open Journal
  of Astrophysics, 2, 8, \dodoi{10.21105/astro.1904.05348}

\bibitem[{{Kaiser} {et~al.}(2022){Kaiser}, {Pol}, {McLaughlin}, {Chen},
  {Hazboun}, {Kelley}, {Simon}, {Taylor}, {Vigeland}, \&
  {Witt}}]{andrew_multiple_gwbs}
{Kaiser}, A.~R., {Pol}, N.~S., {McLaughlin}, M.~A., {et~al.} 2022, arXiv
  e-prints, arXiv:2208.02307.
\newblock \doarXiv{2208.02307}

\bibitem[{{Kelley}(in prep.)}]{holodeck_methods}
{Kelley}, L.~Z. in prep.

\bibitem[{{Kelley} {et~al.}(2017{\natexlab{a}}){Kelley}, {Blecha}, \&
  {Hernquist}}]{Luke_paper1}
{Kelley}, L.~Z., {Blecha}, L., \& {Hernquist}, L. 2017{\natexlab{a}}, \mnras,
  464, 3131, \dodoi{10.1093/mnras/stw2452}

\bibitem[{{Kelley} {et~al.}(2017{\natexlab{b}}){Kelley}, {Blecha}, {Hernquist},
  {Sesana}, \& {Taylor}}]{Luke_paper2}
{Kelley}, L.~Z., {Blecha}, L., {Hernquist}, L., {Sesana}, A., \& {Taylor},
  S.~R. 2017{\natexlab{b}}, \mnras, 471, 4508, \dodoi{10.1093/mnras/stx1638}

\bibitem[{{Kelley} {et~al.}(2018){Kelley}, {Blecha}, {Hernquist}, {Sesana}, \&
  {Taylor}}]{Luke_single_source}
---. 2018, \mnras, 477, 964, \dodoi{10.1093/mnras/sty689}

\bibitem[{Lam {et~al.}(2015)Lam, Pitrou, \& Seibert}]{numba_paper}
Lam, S.~K., Pitrou, A., \& Seibert, S. 2015, in Proceedings of the Second
  Workshop on the LLVM Compiler Infrastructure in HPC, LLVM '15 (New York, NY,
  USA: Association for Computing Machinery), \dodoi{10.1145/2833157.2833162}

\bibitem[{Lam {et~al.}(2022)Lam, stuartarchibald, Pitrou, Florisson, Seibert,
  Markall, esc, Anderson, rjenc29, Leobas, luk-f a, Bourque, Meurer, Collison,
  Oliphant, densmirn, njwhite, Totoni, Pronovost, Seefeld, Grecco, Wieser,
  Peterson, Virshup, G, Turner-Trauring, Bourbeau, Malakhov, Laserson, \&
  Maries}]{numba_zenodo}
Lam, S.~K., stuartarchibald, Pitrou, A., {et~al.} 2022, numba/numba: Version
  0.55.0, 0.55.0,  Zenodo, \dodoi{10.5281/zenodo.5847553}

\bibitem[{Mingarelli {et~al.}(2013)Mingarelli, Sidery, Mandel, \&
  Vecchio}]{Chiara_anisotropy_2013}
Mingarelli, C. M.~F., Sidery, T., Mandel, I., \& Vecchio, A. 2013, Phys. Rev.
  D, 88, 062005, \dodoi{10.1103/PhysRevD.88.062005}

\bibitem[{{Mingarelli} {et~al.}(2017){Mingarelli}, {Lazio}, {Sesana}, {Greene},
  {Ellis}, {Ma}, {Croft}, {Burke-Spolaor}, \&
  {Taylor}}]{chiara_local_gw_landscape}
{Mingarelli}, C. M.~F., {Lazio}, T. J.~W., {Sesana}, A., {et~al.} 2017, Nature
  Astronomy, 1, 886, \dodoi{10.1038/s41550-017-0299-6}

\bibitem[{{Phinney}(2001)}]{Phinney}
{Phinney}, E.~S. 2001, arXiv e-prints, astro.
\newblock \doarXiv{astro-ph/0108028}

\bibitem[{{Pol} {et~al.}(2022){Pol}, {Taylor}, \&
  {Romano}}]{Nihan_anisotropy_forecast}
{Pol}, N., {Taylor}, S.~R., \& {Romano}, J.~D. 2022, arXiv e-prints,
  arXiv:2206.09936.
\newblock \doarXiv{2206.09936}

\bibitem[{{Pol} {et~al.}(2021){Pol}, {Taylor}, {Kelley}, {Vigeland}, {Simon},
  {Chen}, {Arzoumanian}, {Baker}, {B{\'e}csy}, {Brazier}, {Brook},
  {Burke-Spolaor}, {Chatterjee}, {Cordes}, {Cornish}, {Crawford}, {Thankful
  Cromartie}, {Decesar}, {Demorest}, {Dolch}, {Ferrara}, {Fiore}, {Fonseca},
  {Garver-Daniels}, {Good}, {Hazboun}, {Jennings}, {Jones}, {Kaiser}, {Kaplan},
  {Shapiro Key}, {Lam}, {Lazio}, {Luo}, {Lynch}, {Madison}, {McEwen},
  {McLaughlin}, {Mingarelli}, {Ng}, {Nice}, {Pennucci}, {Ransom}, {Ray},
  {Shapiro-Albert}, {Siemens}, {Stairs}, {Stinebring}, {Swiggum}, {Vallisneri},
  {Wahl}, {Witt}, \& {Nanograv Collaboration}}]{Astro4Cast}
{Pol}, N.~S., {Taylor}, S.~R., {Kelley}, L.~Z., {et~al.} 2021, \apjl, 911, L34,
  \dodoi{10.3847/2041-8213/abf2c9}

\bibitem[{{Rajagopal} \& {Romani}(1995)}]{smbh_gwb_rajagopal}
{Rajagopal}, M., \& {Romani}, R.~W. 1995, \apj, 446, 543,
  \dodoi{10.1086/175813}

\bibitem[{{Romano} {et~al.}(2021){Romano}, {Hazboun}, {Siemens}, \&
  {Archibald}}]{romano_crn_vs_gwb}
{Romano}, J.~D., {Hazboun}, J.~S., {Siemens}, X., \& {Archibald}, A.~M. 2021,
  \prd, 103, 063027, \dodoi{10.1103/PhysRevD.103.063027}

\bibitem[{{Rosado} {et~al.}(2015){Rosado}, {Sesana}, \&
  {Gair}}]{rosado_expected_properties}
{Rosado}, P.~A., {Sesana}, A., \& {Gair}, J. 2015, \mnras, 451, 2417,
  \dodoi{10.1093/mnras/stv1098}

\bibitem[{{Sesana} {et~al.}(2004){Sesana}, {Haardt}, {Madau}, \&
  {Volonteri}}]{smbh_gwb_sesana}
{Sesana}, A., {Haardt}, F., {Madau}, P., \& {Volonteri}, M. 2004, \apj, 611,
  623, \dodoi{10.1086/422185}

\bibitem[{{Sesana} {et~al.}(2008){Sesana}, {Vecchio}, \&
  {Colacino}}]{Sesana_high_f_discrepancy}
{Sesana}, A., {Vecchio}, A., \& {Colacino}, C.~N. 2008, \mnras, 390, 192,
  \dodoi{10.1111/j.1365-2966.2008.13682.x}

\bibitem[{{Sesana} {et~al.}(2009){Sesana}, {Vecchio}, \& {Volonteri}}]{SVV09}
{Sesana}, A., {Vecchio}, A., \& {Volonteri}, M. 2009, \mnras, 394, 2255,
  \dodoi{10.1111/j.1365-2966.2009.14499.x}

\bibitem[{{Taylor}(2021)}]{SteveBook}
{Taylor}, S.~R. 2021, arXiv e-prints, arXiv:2105.13270.
\newblock \doarXiv{2105.13270}

\bibitem[{Taylor \& Gair(2013)}]{Steve_anisotropy_2013}
Taylor, S.~R., \& Gair, J.~R. 2013, Phys. Rev. D, 88, 084001,
  \dodoi{10.1103/PhysRevD.88.084001}

\bibitem[{Taylor {et~al.}(2020)Taylor, van Haasteren, \&
  Sesana}]{Steve_bumpy_background}
Taylor, S.~R., van Haasteren, R., \& Sesana, A. 2020, Phys. Rev. D, 102,
  084039, \dodoi{10.1103/PhysRevD.102.084039}

\bibitem[{{Vagnozzi}(2021)}]{inflation_gwb}
{Vagnozzi}, S. 2021, \mnras, 502, L11, \dodoi{10.1093/mnrasl/slaa203}

\bibitem[{{van Haasteren} {et~al.}(2011){van Haasteren}, {Levin}, {Janssen},
  {Lazaridis}, {Kramer}, {Stappers}, {Desvignes}, {Purver}, {Lyne}, {Ferdman},
  {Jessner}, {Cognard}, {Theureau}, {D'Amico}, {Possenti}, {Burgay},
  {Corongiu}, {Hessels}, {Smits}, \& {Verbiest}}]{EPTA_cosmic_string}
{van Haasteren}, R., {Levin}, Y., {Janssen}, G.~H., {et~al.} 2011, \mnras, 414,
  3117, \dodoi{10.1111/j.1365-2966.2011.18613.x}

\bibitem[{{Vogelsberger} {et~al.}(2014){Vogelsberger}, {Genel}, {Springel},
  {Torrey}, {Sijacki}, {Xu}, {Snyder}, {Nelson}, \& {Hernquist}}]{Illustris}
{Vogelsberger}, M., {Genel}, S., {Springel}, V., {et~al.} 2014, \mnras, 444,
  1518, \dodoi{10.1093/mnras/stu1536}

\bibitem[{Zonca {et~al.}(2019)Zonca, Singer, Lenz, Reinecke, Rosset, Hivon, \&
  Gorski}]{healpy}
Zonca, A., Singer, L., Lenz, D., {et~al.} 2019, Journal of Open Source
  Software, 4, 1298, \dodoi{10.21105/joss.01298}

\end{thebibliography}
\bibliographystyle{aasjournal}



\end{document}